\documentclass[preprint,11pt,authoryear]{elsarticle}
\newcommand{\bx}[1]{\mbox{#1}\xspace}
 
\newcommand{\fc}[1]{\ensuremath{\bx{\textsc {\large #1}}}\xspace}

\newcommand{\enc}[2]{\ensuremath{\bx {\large $$\textsc{Enc}$(#1,$ $$#2$)$}}\xspace}
\newcommand{\m}[1]{\ensuremath{\bx{\sffamily {#1}}}\xspace}
\newcommand{\mac}[2]{
\ifthenelse{\equal{#2}{}}
{\ensuremath{\bx {\large $$\textsc{Mac}$_{#1}$}}\xspace}
{\ensuremath{\bx {\large $$\textsc{Mac}$(#1,$ $$#2$)$}}\xspace}
}
\newcommand{\sig}[2]{
\ifthenelse{\equal{#2}{}}
{\ensuremath{\bx {\large $$\textsc{Sign}$_{#1}$}}\xspace}
{\ensuremath{\bx {\large $$\textsc{Sign}$(#1,$ $$#2$)$}}\xspace}
}
\newcommand{\cert}[1]{\ensuremath{$cert$_{\bx{\footnotesize {\em #1}}}}\xspace}
\newcommand{\servlist}{\ensuremath{$`serv[1...n]'$}\xspace}
\newcommand{\serv}{\ensuremath{$`serv[i]'$}\xspace}
\newcommand{\service}{\ensuremath{$serv[i]$}\xspace}
\newcommand{\assert}{\ensuremath{$assert$}\xspace}

\newcommand{\pvt}[1] {\ensuremath{S_{\bx{\footnotesize {\em {#1}}}}}\xspace}
\newcommand{\skey}[3]{\ensuremath{k^{\bx{\footnotesize {\em $#1$}}}_{\bx{\footnotesize {\em #2,#3}}}}\xspace}
\newcommand{\pub}[1]{\ensuremath{P_{\bx{\footnotesize {\em {#1}}}}}\xspace}
\newcommand{\n}[1]{\ensuremath{$n$_{\bx{\footnotesize {\em $#1$}}}}}
\newcommand{\linha}[1]{\ensuremath{$n'$_{\bx{\footnotesize {\em $#1$}}}}}
\newcommand{\dlinha}[1]{\ensuremath{$n''$_{\bx{\footnotesize {\em $#1$}}}}}

\newcommand{\tx}[2]{\ensuremath{#1\rightarrow}#2\xspace}

\usepackage{amssymb}
\usepackage{balance}
\usepackage{comment}
\usepackage{leob}
\usepackage{xcolor}
\usepackage[modulo]{lineno}
\usepackage{float}
\usepackage{multirow}
\usepackage{url}
\usepackage{subfig}
\usepackage{newfloat}
\usepackage{pbox}
\usepackage{tikz}
\usepackage{tabularx}
\usepackage{adjustbox}
\usepackage[margin=3.0cm]{geometry}
\newcolumntype{T}[1]{>{\raggedleft\arraybackslash$}m{#1}<{$}}
\newlength{\mycolwd}                                         
\DeclareFloatingEnvironment[
    fileext=los,
    listname={List of Protocols},
    name=Protocol,
    placement=tbhp,
]{protocol}

\journal{Ad Hoc Networks}

\begin{document}

\begin{frontmatter}



\title{A Federated Lightweight Authentication Protocol for the Internet of Things}



\author[1]{Maria L. B. A. Santos
}
\ead{mburga@dcc.ufmg.br}
\author[1]{J\'essica C. Carneiro}
\ead{jessicarneiro@dcc.ufmg.br}
\author[1]{Ant\^onio M. R. Franco}
\ead{franco@dcc.ufmg.br}
\author[4]{Fernando A.\ Teixeira}
\ead{teixeira@ufsj.edu.br}
\author[5]{Marco A. Henriques}
\ead{marco@dca.fee.unicamp.br}
\author[6]{Leonardo B.\ Oliveira\corref{cor1}%
\fnref{1}}
\ead{leob@dcc.ufmg.br}

\cortext[cor1]{Corresponding author}
\address[1]{UFMG, Belo Horizonte, Brazil}
\address[4]{UFSJ, Ouro Branco, Brazil}
\address[5]{Unicamp, Campinas, Brazil}
\address[6]{Visiting professor at Stanford University, Stanford, USA}

\begin{abstract}
Considering the world's IoT development and market, it is necessary to guarantee the security of the developed IoT applications as well as the privacy of their end users. In this sense, Federated Identity Management (FIdM) systems can be of great help as they improve user authentication and privacy. In this paper, we claim that traditional FIdM are mostly cumbersome and then ill-suited for IoT. As a solution to this problem, we come up with a federated identity authentication protocol exclusively tailored to IoT. Federated Lightweight Authentication of Things (FLAT), our solution, replaces weighty protocols and asymmetric cryptographic primitives used in traditional FIdM by lighter ones. For instance, FLAT synergistically combines symmetric cryptosystems and Implicit Certificates. The results show that FLAT can reduce the data exchange overhead by around 31\% when compared to a baseline solution. FLAT's Client is also more efficient than the baseline solution in terms of data transmitted, data received, total data exchange, and computation time. Our results indicate that FLAT runs efficiently even on top of resource-constrained devices like Arduino.

\end{abstract}




\end{frontmatter}


\section{Introduction}
\label{sec:intro}

The development of the Internet of Things (IoT)~\cite{atzori2010,gubbi2013,borgia2014} is a national priority in several countries around the world and is significantly impacting our society. Studies suggest that we will be surrounded by around 20 billion IoT devices in 2020\footnote{https://www.gartner.com/en/newsroom/press-releases/2017-02-07-gartner-says-8-billion-connected-things-will-be-in-use-in-2017-up-31-percent-from-2016}. The IoT development has enabled a diverse number of applications in academy and industry.  





When it comes to IoT, one of the most significant challenges to its full realization lies in the field of Identity Management (IdM). IdM refers to the identification of users in a given system (e.g., a network, application, or service) and controlling their access to resources within that system. (Here, the term {\em identification} means the process for authenticating the identity of a user~\cite{stallings2016cryptography}.). Ideally, IdM provides administrators with the tools to manage the full user's identity life-cycle (e.g., setup, maintenance, and tear down) and, thus, are vital to the security and productivity of organizations.

\begin{figure}[!h] 
  \centering
  \includegraphics[scale=0.65]{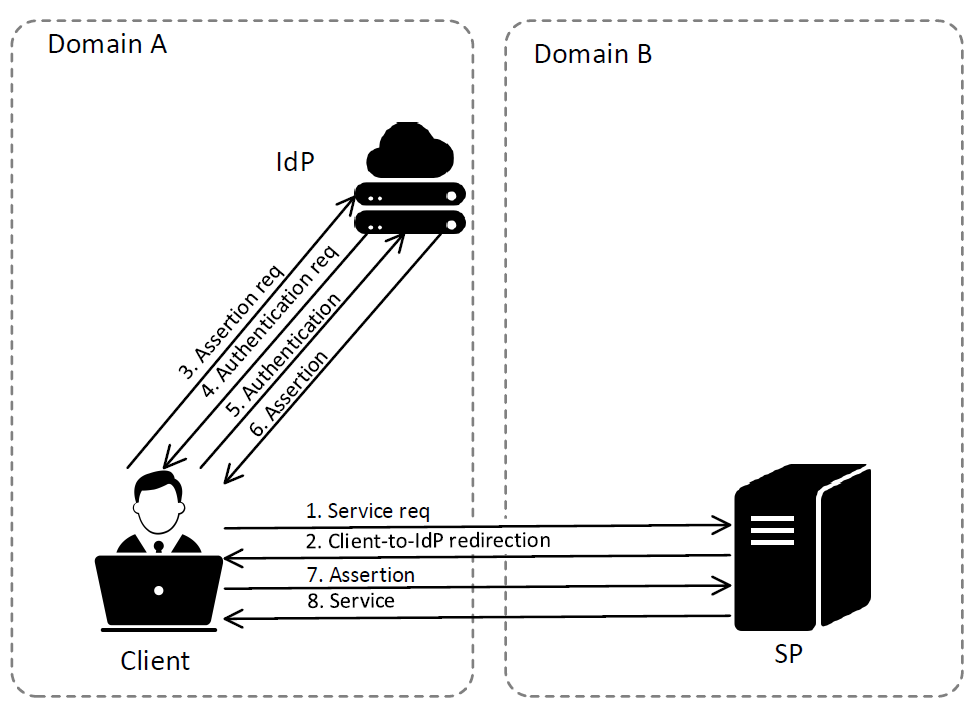}    
  \caption{Traditional FIdM (adapted from \cite{birrell2013}, Figure 2).}
  \label{fig:fidm-trad} 
\end{figure}


Federated Identity Management (FIdM)~\cite{shim2005federated}, in turn, improves the IdM idea by enabling a domain to control accesses to its resources from external users. For instance, it allows an external user to authenticate to a local server and utilize its services without having to create an identity or register credentials locally. Instead, the authentication process between the user and the Service Provider (SP) is mediated by the Identity Provider (IdP) of the user's home domain. FIdM, hence:

\begin{itemize}
\item enables applications like Single-Sign-On (SSO);
\item increases privacy by limiting the amount of information shared;
\item and improves the end-users experience and security by eliminating the need for new accounts registration and restricting the number of entities that hold their password.
\end{itemize}

\strf{fidm-trad} illustrates how traditional FIdM works. The Client initiates the communication with the SP (\strf{fidm-trad}, step 1). The SP, in turn, redirects the Client to the IdP (step 2). Next, the Client requests an assertion (or token) to the IdP (step 3) and authenticates itself to the IdP by presenting its credentials (steps 4 and 5). The IdP, in exchange, sends the assertion to the Client (step 6). This assertion is then used by the Client to get access to the SP service (steps 7 and 8).  All these steps are normally protected and authenticated by public key infrastructure certificates, which have to be validated by all participants.

So far, unfortunately, IoT technology cannot fully enjoy the benefits of either IdM or FIdM. This is so because both IdM or FIdM widely adopted approaches are inadequate to IoT~\cite{fremantle2014}. First and foremost, there is no such thing as IdM for IoT devices. Instead, in existing (F)IdM schemes, IoT devices make use of credentials of individual (human) users to log on and enjoy domain services. And this is both insecure and inappropriate as, respectively, devices most likely should not have the same clearance level as those users and some IoT devices just cannot be naturally linked to any individual user (e.g., whose user should a traffic light be?). Second, the IoT mobile nature and dynamics urge a higher level of scalability and interoperability across multiple domains when compared to conventional network elements~\cite{fremantle2014}. And last but not least, the authentication process on existing (F)IdM schemes normally build upon the login/password paradigm, which is usually okay for humans but ill-suited for devices~\cite{fremantle2016oauthing}, and leverage expensive RSA/DSA cryptosystems, hence, incurring significant computational resources overhead~\cite{perrig02spins}. 
So, there is patently a dire need for a FIdM able to meet IoT special needs.

As a solution to this problem, we propose a FIdM protocol exclusively tailored to IoT. In short, we design, developed, and evaluate a prototype of such a protocol. Our solution, called Federated Lightweight Authentication of Things (FLAT) \cite{santos2018}, replaces cumbersome protocols and asymmetric (public-key) cryptosystems used in traditional FIdM by more efficient ones and, thus, well-suited for IoT \footnote{Santos et al. \cite{santos2018} present the general concept of FLAT and preliminary results of the solution.}. Notably, FLAT synergistically combines symmetric cryptosystems and  Implicit Certificates~\cite{ecqv2001} to come up with a federated authentication protocol. 
We built a prototype of FLAT and evaluated its performance.

This paper is organized as follows. We first argue the need for lightweight cross-domain authentication schemes (\strs{need}). Next, we present how FLAT meet this need (\strs{prot}). We describe FLAT development (\strs{devel}), discuss its performance figures (\strs{eval}), and then sum up our findings (\strs{con}). 

\section{The Call for a Lightweight Cross-domain Authentication}
\label{sec:need}

There are many situations when IoT nodes from different domains need to talk to each other in a secure manner. A real-world situation that came to our attention by way of one of the biggest technology company is this: Suppose that an autonomous (driverless) truck that hauls iron ore has gotten out of order in the middle of a mine\footnote{https://www.technologyreview.com/s/603170/mining-24-hours-a-day-with-robots/}. And that a technical rescue team has to provide {\em in-situ} emergency care in order to get the truck fixed. To do so, ideally, the rescue team needs its technical apparatus to communicate seamlessly with the truck's engine to start providing assistance. Note, here, we have devices (the engine and the technical equipment) from distinct domains (the mining company and the technical rescue company) that need to inter-operate as soon as possible (as this truck broken down may nearly lead to million-dollar losses per day) and in an authenticated manner (because a truck like this is very expensive and cannot be accessed by someone or something it does not recognize).

Another example of such a need is automated cashless toll systems. Here,  cars (clients) are equipped with tags that automatically authenticates to the toll gate, authorizing the car to get into the tollway after processing the payment. Note, however, that a plain implementation of such a system  (i.e., without the employment of FIdM or other security countermeasures to protect privacy) is inappropriate to both car users and tollway companies. The system is inappropriate to the user because it trades privacy for convenience\footnote{https://www.theverge.com/2013/3/27/4150702/golden-gate-bridges-new-cashless-tollway-promises-convenience-for-privacy}. For instance, it makes it possible for the toll company to keep track of the users' mobility pattern and then infer part of users daily routine. And the system is also inappropriate to the toll company because it may not comply with contemporary data privacy laws (\eg~The EU General Data Protection Regulation).


Note the employment of a FIdM system could easily solve the above privacy issue were it not for the resources it requires. For instance, one of the most popular protocol for cross-domain message exchange, the Security Assertion Markup Language (SAML)~\cite{maler2008}, relies on asymmetric cryptography and then is inadequate to the IoT scenario, due to the computational resources required. There are indeed streamlined FIdM proposals, some even tailored to IoT (\eg~\cite{fremantle2016oauthing},~\cite{cirani2015iot},~\cite{domenech2016}). However, those proposals still require asymmetric operations from Clients and, therefore, are not ideal for resource-constrained environments. So, the major challenge for solving the privacy issue lies in designing a FIdM that fits the resources of IoT devices. 


\section{FLAT: A Federated Lightweight Authentication Protocol for IoT}
\label{sec:prot}

FLAT adapts the traditional FIdM authentication to achieve a federated lightweight authentication solution for IoT. In what follows, we talk about the FLAT various assumptions (\strs{assume}), present FLAT itself (\strs{desc}), discuss FLAT design decision making (\strs{high}), and then present FLAT message description (\strs{msg}).

\subsection{Assumptions}
\label{sec:assume}

FLAT focus on a critical part of a FIdM model: authentication. For other parts, FLAT assumes the use of existing approaches. Below, we describe FLAT assumptions.\\

\subsubsection{Device capability}

FLAT assumes the Client, SP, and IdP will be run over devices of low, medium, and high computational capabilities, respectively. This is so because:
\begin{itemize}
\item IoT devices are normally resource-constrained, making the Client the main motivation of our work;
\item the SP, in the end, is still a kind of server and, most likely, will be endowed with more resources than the Client;
\item and the IdP is probably like any other IdPs, \ie a Cloud server that will run 24/7.
\end{itemize}

By definition,  IoT devices are interconnected via the Internet. As for the Client, this can be done either through the devices' own capabilities (\eg by featuring a mobile chip like modern e-book readers/smartwatches usually do) or indirectly by requiring the device to connect to a hotspot close to the SP. After obtaining the SP service broadcast, the Client could also ask the SP to forward its packets to the IdP, as it is expected that the SP is permanently connected to the Internet. 

\subsubsection{Service Discovery}
Service discovery allows participants of a network to do things like service announcement and consultation~\cite{ververidis2008}. In IoT,  devices use services provided by SPs physically close to them and, thus, SPs convey service information using beacons,\ie periodically broadcasted messages within a certain radius. FLAT assumes that the Client discovers the SP and its services using beacons.\\

\subsubsection{Trust Establishment}
FLAT assumes the trust between IdPs and Clients as well as IdPs and SPs takes place a priori. FLAT assumes the trust between IdPs and Clients is established during a Client setup, where shared symmetric keys are pre-loaded into IdPs and Clients. 

As for IdPs and SPs, they are normally from different domains and, previously, not known to each other. FLAT assumes the trust between them derives from digital certificates issued by a common Certification Authority.\\ 

\subsubsection{Interfederation}
In case the Client tries to use a service offered in a different federation, FLAT assumes a model based on the GÉANT Authorisation INfrastructure for the research and education community (eduGAIN)\footnote{https://www.geant.org/Services/Trust\_identity\_and\_security/eduGAIN} for communication between different federations. eduGAIN is an interfederation service that allows the connection of academic identity federations and their respective infrastructures for authentication and authorization. The interfederation can be seen as a way of making certificates issued by one federation Certification Authority valid in another federation. This is done through cross-certification or the creation of a new root Certification Authority that will issue certificates for the Certification Authorities of all participant federations.\\ 

\begin{figure}[!t] 
  \centering
  \includegraphics[scale=0.7]{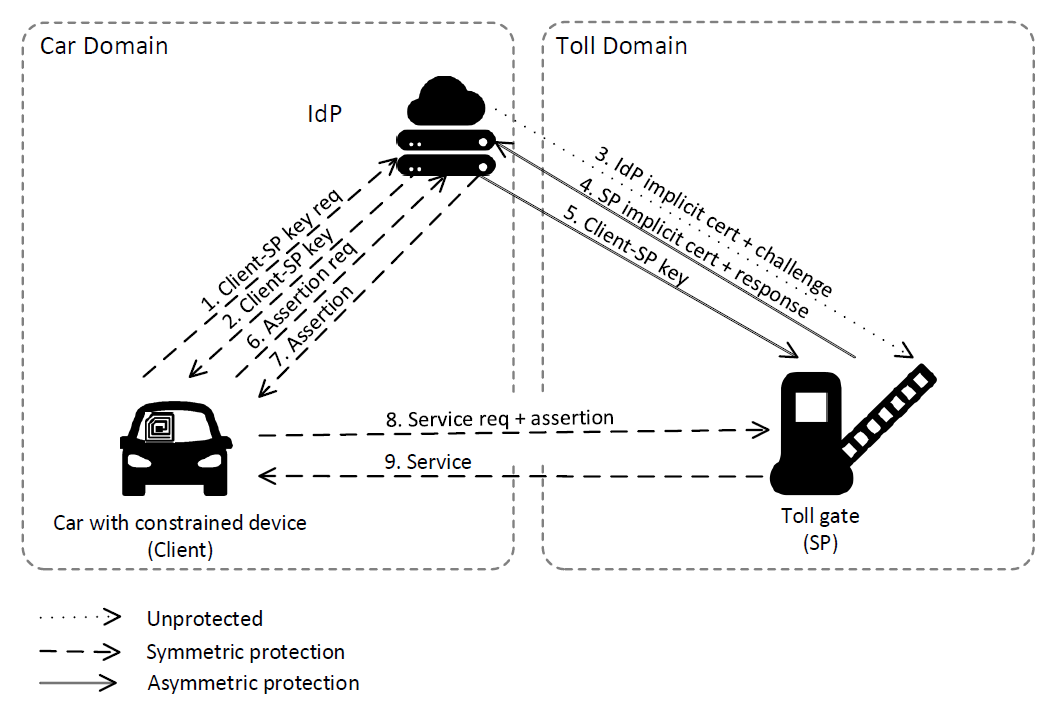}    
  \caption{FLAT applied to a tollway scenario.}
  \label{fig:flat} 
\end{figure}

\subsubsection{Setup}
FLAT assumes that a number of procedures will take place prior to the Clients deployment. A central point for security is the generation of cryptographic keys. Physical Unclonable Functions (PUFs) are known to be resistant to physical attacks. FLAT, therefore, assumes the use of PUFs for generating the device’s cryptographic keys~\cite{suh2007}. In FLAT, briefly, we assume the PUF 
output is generated, formatted and stored in each device, shared with the IdP and then used as a key to secure the Client$\leftrightarrow$IdP communication. We assume this whole process takes place in a secure environment (\eg at a physically secure room inside the device's home domain.)\\ 

\begin{protocol*}[!t]
\begin{adjustbox}{max width=\textwidth}
\setlength\arraycolsep{5pt}
$\begin{array}{r@{}T{\mycolwd}l}
\multicolumn{3}{l}{\fc{Operation($Client$ $C$, $Identity$ $Provider$ $IdP$, $Service$ $Provider$ $SP$)}}                  \\
  1.1. & \tx{C}{IdP}:               & \mac{\skey{}{IdP}{C}}{\n{C} $\mid$ SP $\mid$ \m{key\_req}}                                         \\
  2.1. & \tx{IdP}{SP}:              & \n{IdP} \mid$ $ \cert{IdP}                   \\
  2.2. & \tx{SP}{IdP}:              & \sig{\pvt{SP}}{\cert{SP} $\mid$ \n{IdP} $\mid$ \n{SP} $\mid$ \linha{SP}}                                                                          \\
  2.3. & \tx{IdP}{SP}:              & \sig{\pvt{IdP}}{\enc{\pub{SP}}{\skey{}{SP}{C}} $\mid$ \n{SP} $\mid$ \linha{IdP}}                                 \\
  2.4. & \tx{SP}{IdP}:              & \sig{\pvt{SP}}{\linha{IdP}}                  \\
  1.2. & \tx{IdP}{C}:        	     & \mac{\skey{}{IdP}{C}}{\enc{\skey{}{IdP}{C}}{\skey{}{SP}{C}} $\mid$ \servlist $\mid$ \n{C} $\mid$ \dlinha{IdP}}                                          \\
  3.1. & \tx{C}{IdP}:               & \mac{\skey{}{IdP}{C}}{\enc{\skey{}{IdP}{C}}{\m{assert\_req(\serv)}} $\mid$ \dlinha{IdP} $\mid$ \linha{C}}                    \\
  3.2. & \tx{IdP}{C}:               & \mac{\skey{}{IdP}{C}}{\enc{\skey{}{IdP}{C}}{\assert = $ ${\ensuremath{\bx {\large $$\textsc{Sign}$(\pvt{IdP}, $\serv$ \mid$ $\linha{SP})$}}\xspace}} $\mid$ \linha{C}}                                      \\
  4.1. & \tx{C}{SP}:                & \mac{\skey{}{SP}{C}}{\enc{\skey{}{SP}{C}}{\assert} $\mid$ \dlinha{C} $\mid$ \linha{SP}}                   \\
  4.2. & \tx{SP}{C}:        	     & \mac{\skey{}{SP}{C}}{\enc{\skey{}{SP}{C}}{\service} $\mid$ \dlinha{C}}                               \\

\\
& \multicolumn{2}{@{}l}{$The$$ $$symbols$$ $$denote$:}\\

 &\tx{}{}:           & $ unicast transmission$ \\
 &\mid$ $:           & $ concatenation$ \\
 &\mac{k}{m}:        & $ MAC over $m$ calculated using key $k$ $ \\
 &\skey{}{X}{Y}:     & $ symmetric key shared by X and Y $ \\
 &\n{X}:             & $nonce generated by X$\\
 &\m{key\_req}:      & $ key request label$ \\
 &\m{assert\_req}:   & $ assert request label$ \\
 &\cert{X}:          & $ X's certificate$ \\
 &\pvt{X}:           & $ X's private key $ \\
 &\pub{X}:           & $ X's public key $ \\
 &\sig{k}{m} :       & $signature over message $m$ using key $k$$\\ 
 &\enc{k}{m} :       & $encryption over message $m$ using key $k$$\\
 &\servlist :        & $ list of services $ \\
 &\serv :            & $ description of service $i$ $ \\
 &\service :         & $ service $i$ $ \\
 &\assert :          & $ assertion$ \\
\end{array}$

\end{adjustbox}
\caption{FLAT protocol description.}
\label{prot:protocol}
\end{protocol*}

\subsection{Description} 
\label{sec:desc}

\strf{flat} shows a high-level overview of FLAT`s operation. For readability, we decided not to represent all the cryptographic primitives in~\strf{flat}, individually. There, instead, dotted lines denote no cryptosystem being used, dashed lines denote symmetric cryptosystems (MACs and secret-key encryption), and solid lines, asymmetric cryptosystems (digital signatures and public-key encryption). 

We use a scenario where a car with an attached constrained IoT device interacts with an automated cashless toll system to illustrate FLAT (\strf{flat}). Here, the toll company represents the SP; the device attached to the car, the Client; and the car domain (the car rental company, the car company owner, the department of motor vehicles etc.) the IdP. In context, the service here is the tollway, \ie the passage through the toll road. Broadly, the Client requests to its IdP a session key and an assertion. The IdP, in turn, sends securely the session key to both the Client and SP. Besides, the IdP sends a signed assertion to the Client. Last, the Client uses the session key to secure the communication with the SP and forward the signed assertion to be validated by it in order to allow access to the requested service.

Notably, whenever the Client wants to access a service (\eg go through a toll gate ), it requests the IdP to issue a key to the SP and the Client itself (\strf{flat}, step 1), so the Client and SP can communicate with each other, securely. The IdP respond to this request and issues the key to the Client  (\strf{flat}, step 2) and SP (\strf{flat}, steps 3, 4, and 5). Regarding the SP, this is a three-hand shake process because the SP and IdP first exchange certificates (in this case, Implicit Certificates -- \strf{flat}, steps  3 and 4) to establish a secure channel for the key issuing (\strf{flat}, step 5). Next, as usual in FIdM, the Client requests and receives an assertion (\strf{flat}, step 6 and 7). Last, the Client requests the service by presenting the assertion to the SP (\strf{flat}, step 8), then the SP provides the service (\strf{flat}, step 9) -- \eg by opening the toll gate.

\strp{protocol} shows the complete description of FLAT, with all steps taken in the authentication process. First, Client sends a key request to IdP, indicating the SP providing the service it would like to access (step 1.1, \strp{protocol}). The IdP then exchange certificates with the SP (steps 2.1 and 2.2, \strp{protocol}) and after verifying SP's certificate, sends a symmetric key to the SP, so it will be able to communicate with the Client (step 2.3, \strp{protocol}). The SP sends a signed nonce to the IdP as a confirmation of the reception of the key (step 2.4, \strp{protocol}). Next, the IdP also sends the symmetric key to the Client (step 1.2, \strp{protocol}). The Client can then send an assertion request to the IdP (step 3.1, \strp{protocol}), that responds with the assertion (step 3.2, \strp{protocol}). This assertion signed by the IdP will be forwarded to the SP (step 4.1, \strp{protocol}), that will then be able to provide the service to the Client (step 4.2, \strp{protocol}).

\subsection{Design Highlights} 
\label{sec:high}

\subsubsection{Efficiency}
In FLAT, we have taken some measurements to guarantee the efficiency of the protocol, especially on the Client side. First, we decided that all communication involving the Client --- the most constrained device --- will be carried out by using symmetric cryptosystems. These are more efficient than asymmetric cryptosystems regarding not only computation but also communication (\eg they do not require certificate exchange). To make this come true and, yet, secure the communication between the Client and SP we use the IdP as a Key Distribution Center, or KDC for short.

Second, we tried to decrease the message exchange load on the Client side. By way of example, instead of the Client first contacting the SP (\strf{fidm-trad}, step 1) for only then being redirected to the IdP (\strf{fidm-trad}, step 3), Clients, in FLAT, start by directly contacting the IdP (\strf{flat}, step 1) as soon as it gets the SP service announcement. (Recall we assume Clients get aware of the SP and the services it provides through the use of Service Discovery -- c.f.~\strs{assume}, Service Discovery.) 


Last, FLAT replaces conventional certificates with Implicit Certificates~\cite{ecqv2001} --- also known as Elliptic Curve Qu-Vanstone (ECQV) --- during the SP-IdP communication (\strf{flat}, steps 3 and 4). Implicit Certificates reconstructs public keys from public information and, hence, there is no need of explicitly storing them in certificates. This, in turn, reduces the sizes of the certificates and then saves bandwidth. Further, the certificate verification in this cryptosystem is performed implicitly which, in turn, makes it more efficient, too. (For detailed information on Implicit Certificates, please refer to~\cite{ecqv2001}.).\\ 

\subsubsection{Security}


\textbf{Threat Model} -- We assume that the adversary has access to the messages exchanged in the protocol, i.e., she can eavesdrop, collect, modify and replay messages previously sent in the protocol. We also consider that the attacker might have physical access to the devices. In order to mitigate risks related to the phyhsical access, as mentioned in \strs{assume}, the device's key is generated using PUFs. We also assume there is a secure channel to transmit the symmetric key shared between Client and IdP in the setup phase.






\begin{table}[htb]
\centering
\caption{Countermeasures adopted by FLAT.}
\label{tab:counter}
\begin{tabular}{l|l}
\textbf{Property} & \textbf{Countermeasure}                              \\ \hline
Authenticity      & MACs and digital signatures                          \\ \hline
Confidentiality   & Symmetric and asymmetric encryption                  \\ \hline
Liveness          & Nonces					         \\ \hline
Integrity         & MACs, digital signatures and PUFs                    \\ \hline
Availability      & Authentication between the parties                   \\ \hline
Privacy           & FIdM model: limit the access to identity information
\end{tabular}
\end{table}

FLAT uses (i) Message Authentication Codes (MACs) and digital signatures for providing both authentication and integrity; (ii) ciphers for confidentiality; and (iii) nonces for liveness, \ie for preventing replay-attacks~\cite{stallings2016cryptography}. 
\strt{counter} summarizes the countermeasures adopted by FLAT in order to reduce the possibilities of attacks related to
each of the security properties.



\textbf{Authenticity} -- FLAT uses MACs and
digital signatures to ensure authenticity in its communications.
In Client $\leftrightarrow$ IdP communication, FLAT relies on a pre-shared symmetric key. When the IdP sends
a message to Client using this key, Client knows it is its IdP, since
only the IdP has access to the symmetric key shared through a secure channel in the pre-deployment stage. The same
applies to the case when the Client is sending a message to its IdP. The symmetric key is also used in MACs that protect the messages being exchanged and ensures its authenticity.
Regarding the IdP $\leftrightarrow$ SP communication, it starts with a certificate exchange. As both SP and IdP certificates
are signed by a common CA, they can be verified by both entities. When the SP
receives a message signed by the IdP and verifies it using the certificate’s public key, the SP knows it was really the IdP
who sent the message. Similarly, the IdP can
verify the authenticity of the messages sent by the SP.
Communication between Client and SP also makes use of a symmetric key distributed by the IdP to both SP and Client. The authenticity of messages sent between Client and SP is
ensured by MACs.

SP’s certificate sent in step 2.2, \strp{protocol} is included within the signature. This signature is not necessary, as the certificate bears the CA’s signature. This certificate was included in the signature for the sake
of readability.
FLAT uses implicit certificates (ECQV) to authenticate IdP $\leftrightarrow$ SP communication. The implicit certificate
scheme is considered secure16, i.e., an attacker is not able to successfully falsify an implicit certificate and a private key
that was not issued by a legitimate CA \cite{brown2009}.

\textbf{Confidentiality} -- In order to protect the confidentiality of the messages containing sensitive information, FLAT uses symmetric encryption
in Client $\leftrightarrow$ IdP and Client $\leftrightarrow$ SP communication,
and also asymmetric encryption in IdP $\leftrightarrow$ SP communication. Messages not carrying sensitive
information, such as nonces and certificates, are not encrypted.
FLAT mitigates the possibility of message eavesdropping, since messages containing sensitive information are encrypted,
and thus, even if an attacker is able to intercept one of FLAT’s messages, she will not be able to decrypt its contents.
FLAT also adopts countermeasures to prevent Man-In-The-Middle attacks (MITM), by requiring
proof that they are really whom they claim to be during the communication. In FLAT’s operation, the IdP $\leftrightarrow$ SP
communication is authenticated through the use of digital signatures, with certificate exchange before the shared key
is sent to the SP. Additionally, Client’s identity is confirmed by the IdP and vice-versa,
by the pre-shared symmetric key. Client $\leftrightarrow$ SP communication is also authenticated since
their identities have already been verified by the IdP before the shared symmetric key was sent to each of them.

\textbf{Liveness} -- FLAT uses in its operation randomly
generated nonces to ensure the liveness of the messages being sent. Thus, even in case an attacker is able to intercept a
message and replays it, the attack will not be successful since the liveness of the message can be verified by the nonces. FLAT also uses a nonce to identify the session with the SP. Besides ensuring the liveness of the message,
the nonce $n'\_SP$ (received by the IdP in step 2.2, \strp{protocol}) is sent to the Client (step 3.2, \strp{protocol}), so when the SP
will receive this message (step 4.1, \strp{protocol}) it will know the message refers to the same service request made in the
previous steps.
Although it is an unlikely possibility, if an attacker has
access to a large number of messages and one of the nonces is repeated at some time, she might be able to perform an
attack using an old message. For the sake of readability, nonce encryption is suppressed from the protocol description.

\textbf{Integrity} -- FLAT uses MACs and digital signatures to ensure the integrity of the messages. In Client $\leftrightarrow$ IdP and Client $\leftrightarrow$ SP
communication MACs are used to provide authenticity and integrity to the messages. IdP $\leftrightarrow$ SP communication is protected using digital signatures
to guarantee non-repudiation, authenticity, and integrity to the messages being exchanged. FLAT’s
adopted countermeasure to protect against tampering and MITM attacks to integrity is to use MACs or digital signatures, as explained before.
To protect against physical tampering, FLAT uses
PUFs to generate keys. Thus, even if
a physical tampering will occur in the IoT device, there will be changes in its behavior and the key generation will be
compromised. In this case, it will not be possible to communicate with the IdP and the protocol will not run.

\textbf{Availability} -- Availability ensures that the
system will be available when the users will need it.Note that in FLAT the Service Discovery (\strs{assume}, Service Discovery) precedes any authentication. Therefore, bogus service advertisement messages can be received by the Client. A fake SP doing such advertising will not have a proper certificate. As the protocol runs by, however, the IdP will eventually figure this out (\strf{flat}, steps 3 and 4) and then abort the communication with the source of these advertisements. 
Thus, the IdP will neither provide the SP with a session key nor the Client with a signed assertion for the required service. Therefore, the Client will be protected by its IdP against fake SPs.

\textbf{Privacy} -- FLAT increases privacy by limiting the number of entities that hold the device’s identity
information by only its own IdP, as it adopts the federated model. This model also increases privacy as it allows the devices to
limit the identity information that they would like to share with the SPs.

\begin{figure*}[ht]
    \begin{center}
        \includegraphics[width=\textwidth]{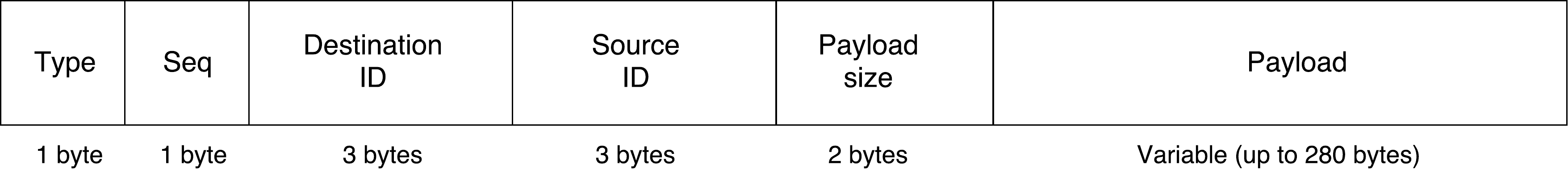}
        \caption{FLAT Message description.}
        \label{fig:msg}
    \end{center}
\end{figure*}

\subsection{\bf Message description.} 
\label{sec:msg}
The message used in FLAT is shown in~\strf{msg}. The first byte is used for message type -- in FLAT, there are ten message types, namely: {\tt key request}, {\tt Client key}, {\tt certificate-challenge}, {\tt certificate-response}, {\tt SP key}, {\tt key acknowledgment}, {\tt assertion request}, {\tt assertion}, {\tt service request}, and {\tt service}. The second byte is used for storing the sequence number. The following two fields have three bytes each and are reserved to store the source and destination identifiers in the application layer. Hence, FLAT supports domains up to $2^{24}$ ($\approx$ 16 million) clients in size. Next, we have a two-byte field to store the payload size. Recipients use this field to get to know the amount of data they should read. Last, we set a payload of at most 280 bytes so that FLAT can accommodate data and cryptographic elements typical of an IoT secure communication protocol.

\section{Development}\label{sec:devel}

We developed a prototype of FLAT. The prototype comprises three modules: Client, SP, and IdP. Both our Client and SP modules are efficient enough to be run on top of resource-constrained IoT devices like Arduino. 
The IdP, conversely, is supposed to be hosted on a server-like device. This is so because we envision the IdP being the single identity provider for all IoT devices in a given domain.

 \begin{figure}[!h]
     \begin{center}
         \includegraphics[scale=0.3]{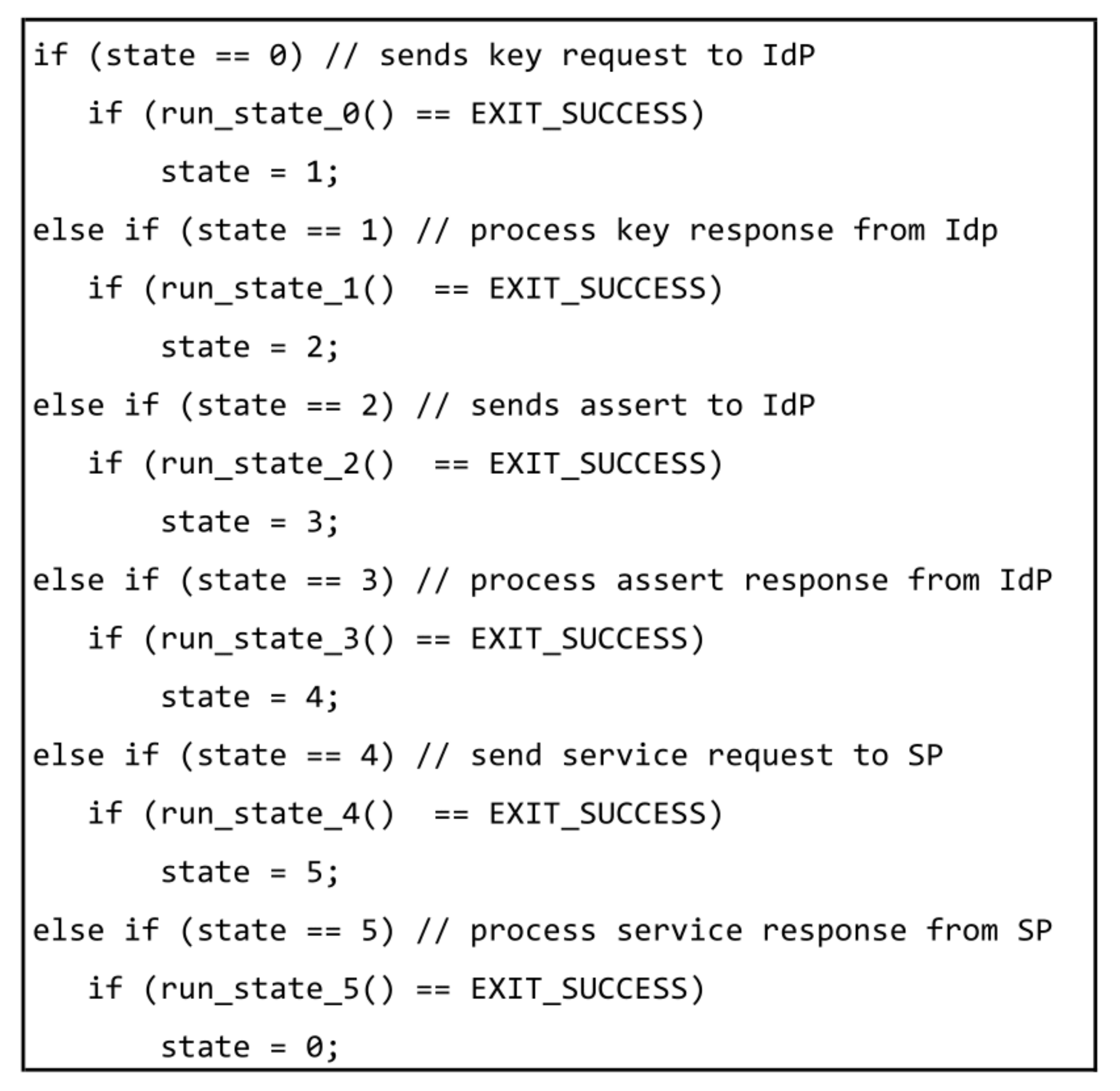}
         \caption{Client's finite-state machine-like implementation.}
         \label{fig:code}
     \end{center}
 \end{figure}

Broadly, our development involved four stages: (i) the development of the communication functions, (ii) the implementation of the cryptosystems, (iii) the integration of the communication functions and the cryptosystems, (iv) the port of the integration to Arduino. 


\subsection{Communication} The communication functions have been developed using the Arduino Wi-Fi library in C/C++. When an Arduino Wi-Fi Shield is attached to an Arduino board, this library makes message transmission using Wi-Fi (802.11b/g) possible. The library provides both UDP and TCP protocols. FLAT was built on top of UDP and was developed based on the idea of a finite-state machine (\strf{code}) going through each step of the authentication protocol. If the synchronization between any pair of interlocutors is lost during the protocol, the process can be restarted through the transmission of a special message.

\begin{figure*}[ht]
    \begin{center}
        \includegraphics[scale=0.12]{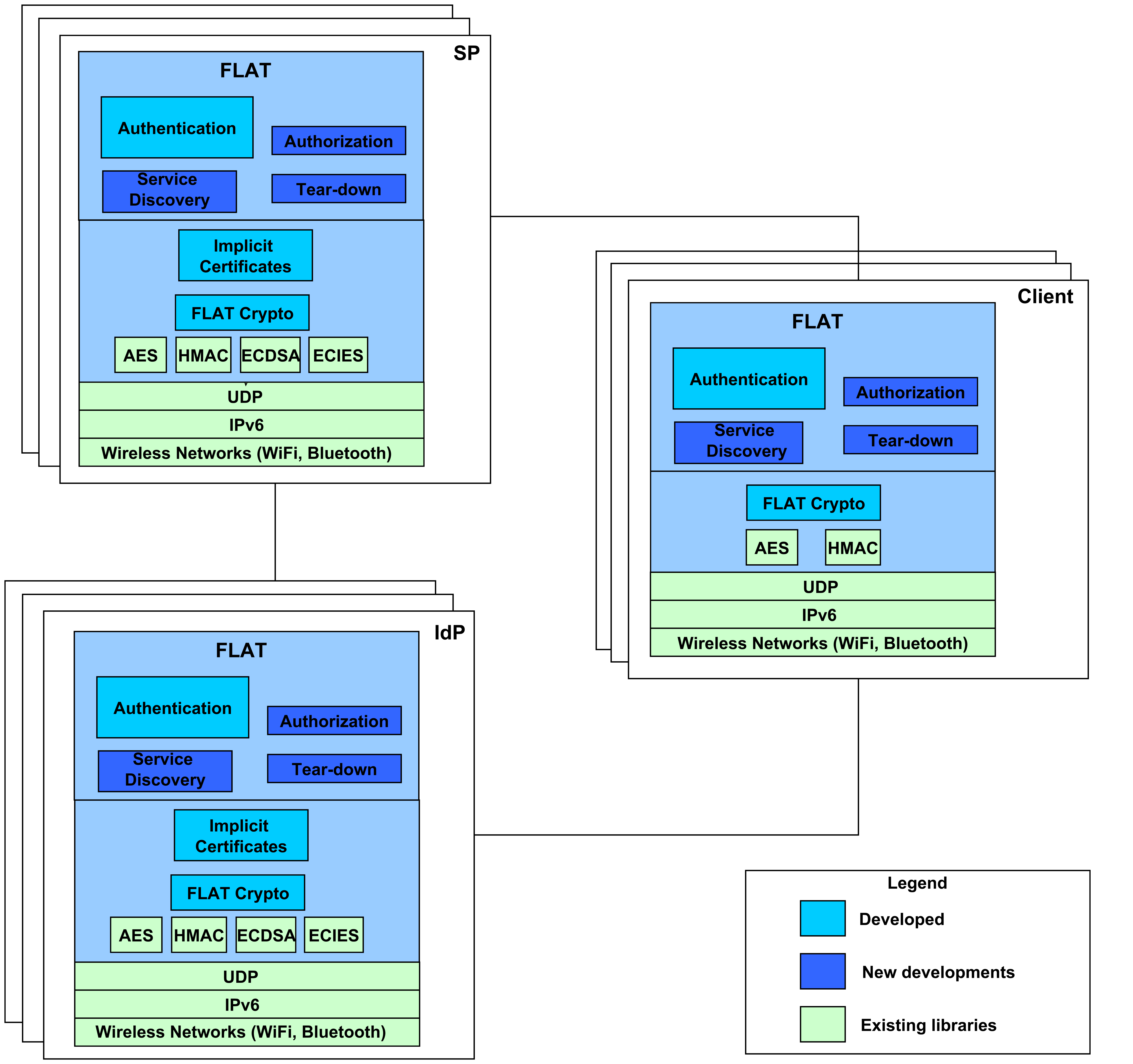}
        \caption{FLAT architecture.}
        \label{fig:archFLAT}
    \end{center}
\end{figure*}


\subsection{Cryptography} All asymmetric cryptosystems used in FLAT are based on Elliptic Curve Cryptography (ECC), as they incur less computation, communication, and storage overhead and therefore suit better the IoT world~\cite{stallings2016cryptography}. Precisely, we use Elliptic Curve Digital Signatures (ECDSA) and the Elliptic Curve Integrated Encryption Scheme (ECIES) as opposed to their traditional counterparts DSA and RSA~\cite{stallings2016cryptography}. 

 We used RELIC\footnote{https://github.com/relic-toolkit} as an underlying cryptographic library. On top of RELIC, we implemented higher-level cryptosystems like Implicit Certificates. Besides, the toolkit already implemented more common cryptosystems needed in FLAT like ECDSA, ECIES, AES, and HMAC. Respectively, FLAT employs these cryptosystems for digitally signing, asymmetrically encrypting, symmetrically encrypting, and message authentication.

\subsection{Security level} Our development of FLAT offers, approximately, a 128-bit security level. In the symmetric part, we have AES using 128-bit keys and HMAC using the hash SHA-256 as a cryptographic hash function. In the asymmetric part---here, all based on ECC---we have the BN-254 elliptic curve as the foundation of the cryptosystems\footnote{Until recently, this curve was thought to offer a 128-bit security level; it offers a 100-bit, though~\cite{barbulescu2017}. The curve can be changed to a more secure one.}.

\subsection{Architecture} \strf{archFLAT} shows the architecture of FLAT, \ie the modules (Client, SP, and IdP) and submodules (communication and cryptography) that comprise our solution. On the Client module, there are only symmetric cryptosystems (AES and HMAC). On the SP and IdP modules, besides the symmetric components, there are also asymmetric cryptosystems (Implicit Certificates, ECDSA, and ECIES).

\begin{figure}[!ht] 
  \centering
\subfloat[]{\label{fig:demo-grant}\includegraphics[width=0.35\columnwidth]{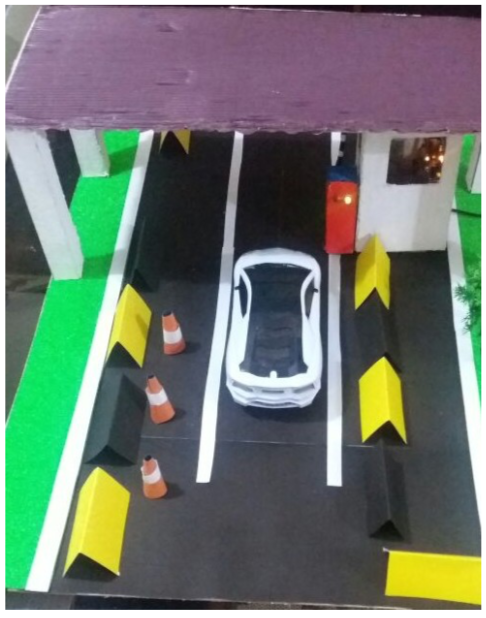}}
 \hfill  
\subfloat[]{\label{fig:demo-den}\includegraphics[width=0.35\columnwidth]{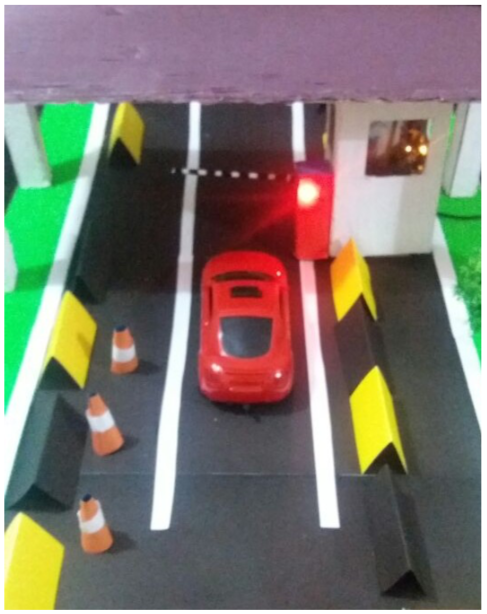}}
  \caption{FLAT demo: access (a) granted and (b) denied.}
  \label{fig:demo}
 \hfill  
\end{figure}

\subsection{Demonstration}\label{sec:demo} 

We developed a demo to showcase our FLAT prototype by demonstrating the functionalities of our architecture and executing operations on multiple classes of devices. Our demo covers the authentication part of FLAT applied to a toy cashless tollway similar to that described in \strs{need}. We say toy because we run FLAT over a mock-up tollway comprised of a toy car and toy gate. (\strf{demo}).

In our demo, a car (client) runs FLAT and tries to authenticate to a toll gate (SP) by using an assertion supplied by its own domain. If the authentication is successful, a green LED lights, the toll gate opens, and the car is granted access to the tollway. Otherwise, red LED lights, the gate keeps closed, and the car's access to the tollway is denied. 


The Client, SP, and IdP have been respectively instantiated by an Arduino Due (84 MHz Atmel, 96 KB SRAM, 512 KB flash), a Raspberry Pi (Raspberry Pi zero W ARM1176JZF-S 1GHz single-core, 512MB RAM), and a Laptop Dell Inspiron (Intel Core i7 2.7 GHz, 8 GB RAM). They correspond to our vision that roles in FLAT will be played by heterogeneous devices (c.f.~\strs{assume}) . For communication, the Arduino board is equipped with an Arduino Wi-Fi shield (802.11b/g), and a MikroTik router plays the role of an access point. For controlling the toll gate, we used a Tower Pro servo motor 9G SG90. Last, we employ some LED bulbs to signal access granted or denied.


\section{Results}\label{sec:eval}

We have evaluated FLAT authentication protocol in terms of computation, communication, protocol total run-time, SRAM and storage, and scalability. For simplicity, we will refer to the authentication protocol as FLAT in this section. Here, our performance numbers correspond to the average value of 100 runs.

Unless otherwise noticed, the numbers for the Client, SP, and IdP in this section refer to numbers measured by running FLAT on top of the demo hardware configuration, namely: Arduino Due, Raspberry Pi, and Laptop Dell Inspiron -- c.f.~\strs{demo} for details. 

\begin{figure}[!ht]
    \begin{center}
        \includegraphics[scale=0.9]{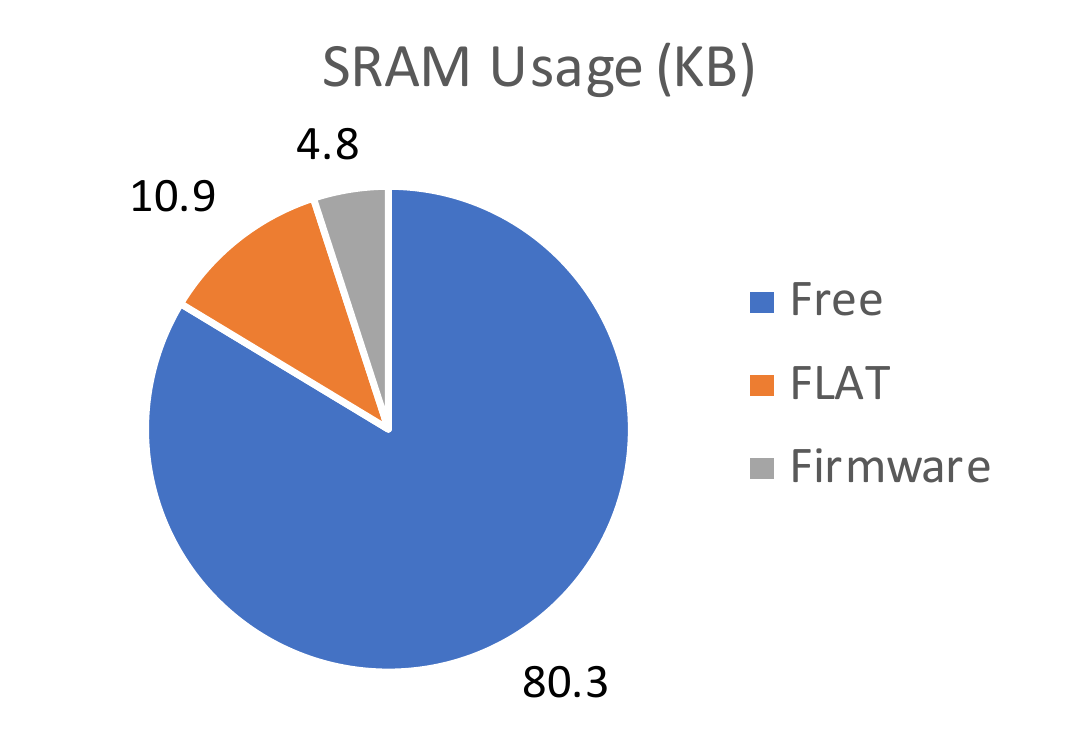}
        \caption{SRAM costs.}
        \label{fig:sram}
    \end{center}
\end{figure}

\subsection{RAM \& Storage} 

We make use of another protocol as a starting point for some comparisons and call it {\em baseline}. The baseline is similar to a traditional FIdM protocol (\strf{fidm-trad}) and does not use Implicit Certificates as FLAT does. For fairness, however, we instantiate baseline using the same parameters as FLAT: certificate fields, level of security, message format, ECC-based cryptosystems, and so forth. 
Baseline certificate size is 134 bytes, as opposed to the Implicit Certificate size used in FLAT which is 70 bytes. This is so because baseline certificate includes, besides the identification information (in our case, 37 bytes), the public key (32 bytes) and the Certificate Authority signature (65 bytes). On the other hand, Implicit Certificates require only the identification information (37 bytes) and the public key extraction data (33 bytes).

We have evaluated FLAT Client requirements regarding SRAM (\strf{sram}) and storage. The results show that FLAT, including the needed Arduino libraries to run the protocol, takes around 16\% of the Client's total memory (Arduino, 15.7 KB out of 96 KB of SRAM). As for storage, FLAT requires 125.8 KB, \ie around 24\% of the Client's total storage (Arduino, 512 KB of flash), approximately. Notably, Arduino's native libraries take 73.5 KB, FLAT communication functions takes 6.8 KB, and cryptography takes 45.5 KB.


\begin{figure}[ht]
    \begin{center}
        \includegraphics[scale=0.9]{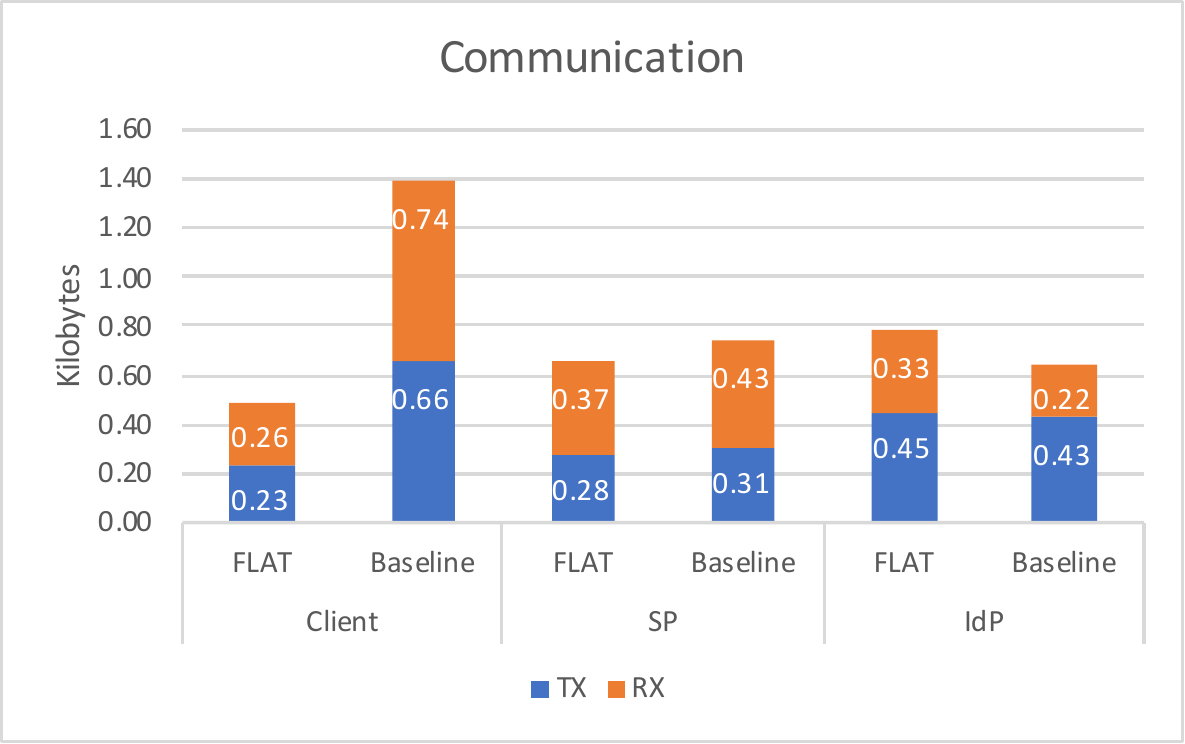}
        \caption{Communication costs.}
        \label{fig:comm}
    \end{center}
\end{figure}

\subsection{Communication}
\label{sec:comm}

\strf{comm} shows the amount of data sent (Tx) and received (Rx) by the Client, SP, and IdP in both FLAT and the baseline solution. In total, FLAT exchanges around 500 bytes on the Client side. It is worth pointing out that the FLAT Client is around 65\% more efficient regarding Tx, Rx, and total data exchanged. Likewise, FLAT, as a whole, is 31\% more efficient than the baseline regarding data exchange (Tx plus Rx). This lower communication overhead on the Client side is mainly because we use symmetric cryptosystems -- they do not require certificate exchange. FLAT SP is around 9\%, 14\%, and 12\%, respectively, more efficient than baseline SP, too. These light Client and SP, however, come to a price. To alleviate the Client and SP, FLAT demands more from the IdP. And then, in FLAT, even using (shorter) Implicit Certificates, the IdP total message exchange is around 20\% higher than in the baseline. We claim, however, the IdP is a powerful device and, therefore, able to handle the total load.

\begin{figure}[ht]
    \begin{center}
        \includegraphics[scale=0.9]{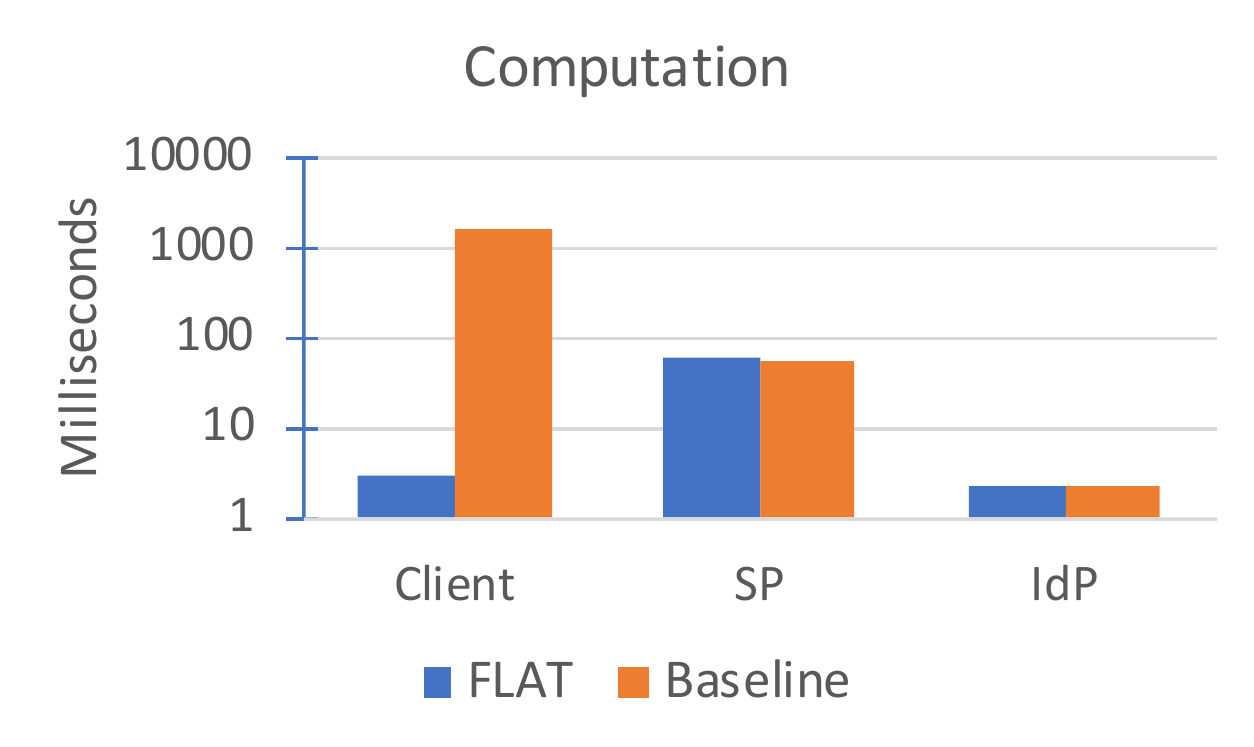}
        \caption{Computation costs.}
        \label{fig:comp}
    \end{center}
\end{figure}

\subsection{Computation}
\label{sec:comp}

FLAT Client is around 520 times faster than baseline (\strf{comp}, Client).
There are two main reasons for that. First, FLAT runs solely symmetric cryptosystems on the Client side. In the baseline, on the other hand, the Client needs to compute the following asymmetric operations: ECIES encryption (1 time), ECIES decryption (1), ECDSA signature (2) and ECDSA signature verification (5). And while symmetric operations are known to be negligible~\cite{perrig02spins}---even in resource-constrained devices---, asymmetric operations---even ECC-based---are known to be inefficient~\cite{malan08}. Second, the Client delegates several tasks to the SP and IdP. A disadvantage of the second reason, however, is that the FLAT SP and IdP end up being  4\% and 2\% less computationally efficient than baseline, respectively. But this is a very small price to pay compared to the gains obtained on other fronts.



\begin{figure}[ht]
    \begin{center}
        \includegraphics[scale=0.9]{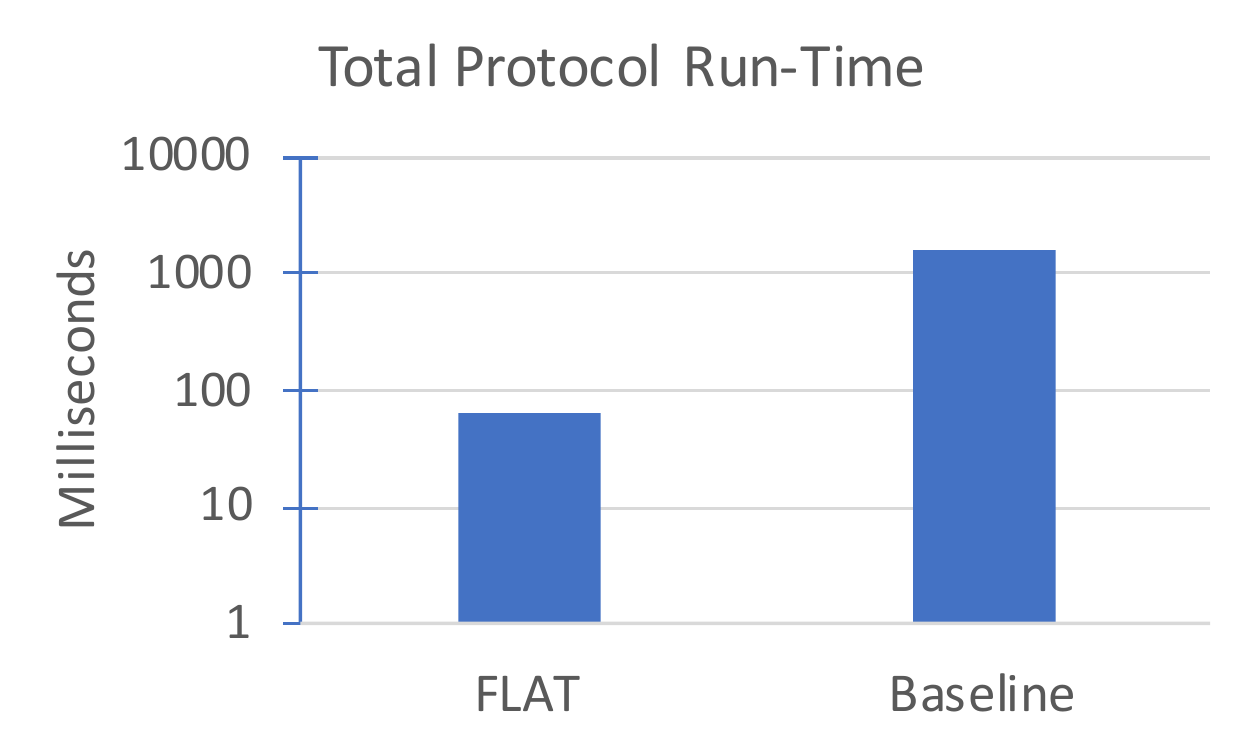}
        \caption{Total protocol run-time.}
        \label{fig:run-time}
    \end{center}
\end{figure}

\subsection{Total Run-time}
\label{sec:run}

We have measured the total amount of time that takes to run FLAT---\ie around 65 ms (\strf{run-time})---and compared that to the baseline solution. Our results show that FLAT total run-time is, on average, only a fraction (around 4\%) of the baseline total run-time (\strf{run-time}). This is a direct consequence of the computation (\strs{comp}) and communication (\strs{comm}) numbers previously shown, \ie FLAT exchanges fewer and smaller messages as well as requires significantly less computational work from the Client.

Notably, the fact that FLAT is computationally more efficient in the slow part of the protocol (Client -- \strs{comp}) --- and only slightly less (4\% and 2\%) in the fast part (SP and IdP, respectively) --- helps in improving its total run-time speedup compared to the baseline solution.

\subsection{Scalability}
\label{sec:scale}


Besides, FLAT scales smoothly regarding the number of Clients. If the number of Clients gets too large for a given IdP, we only need to create one more IdP---certified by the same Certificate Authority---and redistribute the Clients among them, evenly. Of course, this could be done in any federation. However, in FLAT, this process is very simple, as we can adopt a limit on the number of clients that an IdP can attend and, if such limit is reached, another IdP would be installed. New clients would be communicating with this new IdP, without any changes to the previous configuration.


\section{Related Work}\label{sec:rw}






FIdM is already a consolidated solution in the traditional Internet, with several standards and protocols being widely used, such as Shibboleth, OpenID and OAuth. Several works address the authentication and authorization issue for the traditional Internet (e.g. \cite{steiner1988, shim2005federated,maler2008,chadwick2009a, chadwick2009b,birrell2013,isaakidis2016,miettinen2016,zuo2017}).

However, considering the IoT scenario, there is still no consensus on how to provide authentication and authorization to users and devices. In this sense, different IdM solutions for IoT have been proposed (\cite{wang2014, xi2016, neto2016, kim2017, li2018}).

Wang et al. \cite{wang2014} propose a two-factor authentication scheme for wireless sensor networks that can provide privacy protection to users. They analyze case studies to show that previous approaches to the privacy issue in wireless sensor networks are not able to fully address the problem. They investigate possible solutions, concluding that public key cryptography is a suitable approach.

Xi et al. \cite{xi2016} present a key agreement and authentication protocol for IoT devices named TDS (The Dancing Signals). The designed solution provides faster key generation when compared to other solutions, and is resistant to predictable channel attacks. TDS is modeled to only provide keys to physically close devices.

Neto et al. \cite{neto2016} propose an authentication and access control solution for each step in the life-cycle of IoT devices. Using attribute and identity-based cryptography, they developed several protocols to enable Attribute-based access control (ABAC) and authentication in IoT environments. The proposed solution can be run in both powerful and constrained devices.

Hammi et al. \cite{hammi2018} present an authentication solution for IoT named "Bubbles of Trust". Their approach is to provide authentication using a decentralized system based on blockchain in order to ensure availability and integrity. Their solution is evaluated in an implementation using Ethereum blockchain.

Zeng et al. \cite{zeng2018} propose a multi-server approach to provide anonymous user authentication for IoT devices. They also performed a security evaluation of the solution.

Kumari et al. \cite{kumari2018} present an ECC-based authentication protocol targeted to IoT and cloud servers. Their solution is evaluated in terms of performance and its security is formally analyzed using the AVISPA tool. 

Ometov et al. \cite{ometov2019} provides a comparison between different multi-factor authentication solutions for IoT, listing its advantages and disadvantages.

Some authentication and authorization proposals for IoT are based on (F)IdM solutions for the traditional Internet \cite{liu2012, fremantle2014, cirani2015iot, fremantle2016oauthing,domenech2016}). The main advantage of such solutions is that they allow integration with traditional FIdM systems while trying to reduce the costs of these solutions by changing some aspects of the original approach.

Liu et al. \cite{liu2012} design an architecture for authentication and authorization in IoT. The authors briefly discuss existing IdM approaches and opt for ECC-based PKC with OpenID for the authentication and RBAC for authorization. The authors also provide a brief security analysis of the proposed solution. 

Fremantle et al. \cite{fremantle2014} propose a federated identity model to provide access control to IoT devices, evaluating the possibility of an authentication and authorization model for IoT based on OAuth 2.0 with the MQTT protocol. In another work, Fremantle et al. \citep{fremantle2016oauthing} propose an extension to the OAuth 2.0 protocol to provide a FIdM solution for IoT devices, as well as enhance the privacy of users. Their solution is called OAuthing and enables users and devices to be registered automatically, providing each user/device anonymous identities. 

Cirani et al. \cite{cirani2015iot} propose an authorization architecture for IoT using an external authorization service based on OAuth, called IoT-OAS. Their architecture transfers the authorization logic from the constrained device to the external service, allowing the device or smart object to protect its resources by only making a request to the authorization framework. 

Domenech et al. \cite{domenech2016} describe an authentication and authorization model for IoT based on SAML and XACML standards, the AAI4WoT. The authors provide a prototype as well as an evaluation in terms of storage, communication, CPU usage, and power consumption. 

These works develop solutions based on traditional FIdM approaches. FLAT, however, is not based on existing FIdM solutions for the traditional Internet, being exclusively tailored to IoT.

There are solutions which approach the authentication and authorization issue in IoT through asymmetric cryptography \cite{oliveira2009, yavuz2013, porambage2014}. 

Oliveira et al. \cite{oliveira2009} propose a solution to provide end-to-end authentication between a sensor node and multiple users. They claim that given the number of applications that a sensor can be connected, the use of digital signatures is the most suitable option to provide end-to-end authentication. They evaluated both ECDSA and BLS signature algorithms in constrained devices, showing the feasibility of the proposed solution.

Yavuz \cite{yavuz2013} proposes a new signature scheme for authentication of resource-constrained devices called ETA (Efficient and Tiny Authentication). The author claims that traditional signature schemes are not suitable for constrained devices as they require high computational power and energy consumption. ETA also provides smaller signatures and key sizes than other existing similar solutions.

Porambage et al. \cite{porambage2014} also address the end-to-end authentication between users and sensors in IoT. The authors present an authentication protocol that takes into consideration the heterogeneity of the network. Their approach uses asymmetric cryptography for authentication through the use of implicit certificates, in order to make the solution more suitable for resource-constrained devices.





These solutions, however, are based on asymmetric cryptography. FLAT, on the other hand, considers an  IoT-tailored protocol for authentication, that makes use of only symmetric cryptographic primitives in the Client device.


\begin{table*}[t]
\centering
\caption{Related work comparison.}
\label{tab:comrw}
\begin{adjustbox}{max width=\textwidth}
\begin{tabular}{l|l|l|l|l|l|l}
\textbf{Work}  & \textbf{\pbox{30cm}{Symmetric Crypto \\on Client}} & \textbf{\pbox{30cm}{Gateway}} & \textbf{\pbox{30cm}{Client \\device}}  & \textbf{Context} & \textbf{Authorization} & \textbf{Authentication} \\ \hline
Oliveira et al.            & --- &--- & Constrained & General & --- & Yes \\ \hline
Horrow et al. 2012    & ---             & Yes       & Constrained      & General          & Yes      & Yes        \\ \hline
Liu et al. 2012	     & ---             & Yes       & ---             & General          & Yes      & Yes        \\ \hline
Hummen et al. 2013    & ---             & Yes       & Constrained      & General          & ---      & Yes        \\ \hline
Yavuz 2013    & ---             & ---       & Constrained      & General          & ---      & Yes        \\ \hline
Porambage 2014 & ---             & ---       & Constrained     & General          & ---      & Yes        \\ \hline
Fremantle et al. 2014 & ---             & Yes       & Constrained      & General          & Yes      & Yes        \\ \hline
Wang et al. 2014 & ---             & ---      & Constrained      & General          & ---      & Yes        \\ \hline
Markmann et al. 2015 & ---             & Yes       & Constrained      & General          & ---      & Yes        \\ \hline
Cirani et al. 2015    & ---             & Yes       & Constrained      & General          & Yes      & ---        \\ \hline
Witkovski et al. 2015 & Yes             & Yes       & Constrained      & Smart home      & Yes      & Yes        \\ \hline
Fremantle et al. 2016 & ---             & Yes       & Constrained      & General          & Yes      & Yes        \\ \hline
Domenech et al. 2016  & ---             & Yes       & Intermediary    & General          & Yes      & Yes        \\ \hline
Hong et al. 2016      & ---             & Yes       & Constrained      & Smart home      & Yes      & ---        \\ \hline
Xi et al.  2016 & ---             & Yes       & Constrained      & Sensors      & ---      & Yes        \\ \hline
Neto et al. 2016    & ---             & ---       & Constrained     & General      & Yes      & Yes        \\ \hline
FLAT                 & Yes             & ---       & Constrained      & General          & ---      & Yes        \\ \hline
\end{tabular}
\end{adjustbox}
\end{table*}

Other works try to transfer the authentication and authorization tasks to an external gateway or to a more powerful node in the network, in order to preserve the resource-constrained IoT devices \cite{horrow12, hummen2013towards, turkanovic2014, markmann2015, kimesiot2017}.

Horrow and Sardana~\cite{horrow12} present the requirements for an IdM system for IoT and propose an IdM framework for IoT using cloud-based technologies. Instead of having computing nodes to process the information sent by sensors, the computation is performed in the cloud. The authors present a general architecture to approach the identity management issue in IoT, but they do not describe in details how would be the protocols to implement this architecture.  


Hummen et al. \cite{hummen2013towards} propose some optimizations to DTLS in order to use certificates in IoT devices, such as pre-validation of certificates in a gateway and a delegation procedure that allows another entity to validate the certificate instead of the constrained device. However, they present preliminary results and only evaluate the computational overhead, not addressing other aspects of the solution.



In a short work, Markmann et al. \cite{markmann2015} approach the authentication for resource-constrained devices in IoT using Identity-Based Cryptography (IBC) and ECC. They assume the presence of a node with more computational power in each domain in order to provide the end-to-end authentication utilizing IBC. This node works as a trusted authority (similarly to a CA), and a federation of these trusted authorities is required in order to provide authentication between different domains.


The solutions proposed by Horrow and Sardana, Hummen et al., and Markmann et al. make use of another entity to perform authentication and authorization tasks. FLAT, however, do not transfer the computation to an external entity, developing a protocol that can be executed even by restricted devices.


There are also works that address FIdM solutions for specific IoT scenarios \cite{he2015, witkovski2015, Hong16Beetle}.


Witkovski et al. \cite{witkovski2015} propose a solution for IoT device authentication based on session keys and symmetric cryptography with the use of a gateway to integrate the traditional Internet and IoT in the maintenance of electronic devices in a smart home scenario. 


Hong et al. \cite{Hong16Beetle} claim that Bluetooth Low Energy (BLE) can be used as an access control alternative in smart homes. A prototype of the BLE access control solution was built on top of resource-constrained devices and applied to specific smart home scenarios. 

Differently from these approaches (targeted to very specific scenarios), FLAT is a broad solution that can be applied to different IoT scenarios. 




Several works propose FIdM solutions for IoT. These solutions, however, do not address completely the IoT IdM issues, as they are based on IdM solutions for the traditional Internet, rely on asymmetric cryptography or make use of external entities to perform computational tasks. FLAT is a federated authentication protocol that can be used to implement a FIdM system even in restricted devices, as it is based on only symmetric cryptographic primitives in the Client side.


\section{Conclusion}
\label{sec:con}

As IoT gains more visibility and IoT solutions are being incorporated in the most diverse fields, there is a real
need to ensure the security of the developed IoT applications and preserve the privacy of their users. 

In this work, we proposed a federated identity authentication protocol exclusively tailored to IoT called FLAT. FLAT considers that Clients will run over devices of low computational capabilities and, thus, requires only symmetric (light) cryptographic operations from them. Besides, FLAT employs Implicit Certificates rather than regular certificates in the IdP$\leftrightarrow$SP communication for saving both computation and bandwidth. FLAT also replaces inefficient cryptosystems like RSA/DSA by others more suitable to IoT and reduces the message load in the Client device. 

This paper presented the assumptions necessary to the correct operation of FLAT, a prototype of the proposed solution applied to a cashless toll system and an evaluation of FLAT in terms of storage, SRAM, communication, computation, and total run-time.

FLAT computation in the Client is 520 times more efficient than the baseline solution. FLAT uses around 16\% of Client’s SRAM and 24\% of Client’s flash memory when running in an Arduino Due device.  Moreover, FLAT total run-time represents 4\% of the total time necessary to run Baseline. Regarding the communication costs, FLAT is also 31\% more efficient than the Baseline protocol when considering the total data exchange. Our results confirm that FLAT can run efficiently on top of IoT devices. 

FLAT is a step towards the development of a complete IdM system that could take care of all the lifecycle of an IoT device, as it provides an authentication solution that is lightweight, contemplates the mobility aspect of devices considering different security domains and has an operational prototype.

As future works, we consider the development of a complete
IdM system. By now, FLAT is an authentication solution for IoT. The next steps are completing the IdM solution with the implementations of the authorization, service discovery process, and tear-down.

Another future work includes the proposal and evaluation of different communication possibilities for the Client device. The Client could use the SP as an intermediate to communicate with the IdP, using SP's permanent Internet connection and enabling the Client to only have a low power Wi-Fi connection. In another possible scenario, the Client will not need a network connection or to communicate directly with the IdP, with the SP resending the messages between Client and IdP.

\bibliographystyle{elsarticle-harv}
\bibliography{bib.bib}

\begin{thebibliography}{46}
\expandafter\ifx\csname natexlab\endcsname\relax\def\natexlab#1{#1}\fi
\providecommand{\url}[1]{\texttt{#1}}
\providecommand{\href}[2]{#2}
\providecommand{\path}[1]{#1}
\providecommand{\DOIprefix}{doi:}
\providecommand{\ArXivprefix}{arXiv:}
\providecommand{\URLprefix}{URL: }
\providecommand{\Pubmedprefix}{pmid:}
\providecommand{\doi}[1]{\href{http://dx.doi.org/#1}{\path{#1}}}
\providecommand{\Pubmed}[1]{\href{pmid:#1}{\path{#1}}}
\providecommand{\bibinfo}[2]{#2}
\ifx\xfnm\relax \def\xfnm[#1]{\unskip,\space#1}\fi
\bibitem[{Atzori et~al.(2010)Atzori, Iera and Morabito}]{atzori2010}
\bibinfo{author}{Atzori, L.}, \bibinfo{author}{Iera, A.},
  \bibinfo{author}{Morabito, G.}, \bibinfo{year}{2010}.
\newblock \bibinfo{title}{The internet of things: A survey}.
\newblock \bibinfo{journal}{Computer networks} \bibinfo{volume}{54},
  \bibinfo{pages}{2787--2805}.
\bibitem[{B.~Oliveira et~al.(2009)B.~Oliveira, Kansal, Priyantha, Goraczko and
  Zhao}]{oliveira2009}
\bibinfo{author}{B.~Oliveira, L.}, \bibinfo{author}{Kansal, A.},
  \bibinfo{author}{Priyantha, B.}, \bibinfo{author}{Goraczko, M.},
  \bibinfo{author}{Zhao, F.}, \bibinfo{year}{2009}.
\newblock \bibinfo{title}{{Secure-TWS: Authenticating Node to Multi-user
  Communication in Shared Sensor Networks}}, in: \bibinfo{booktitle}{ACM
  International Conference on Information Processing in Sensor Networks
  (IPSN'09)}, \bibinfo{organization}{ACM}. pp. \bibinfo{pages}{289--300}.
\bibitem[{Barbulescu and Duquesne(2017)}]{barbulescu2017}
\bibinfo{author}{Barbulescu, R.}, \bibinfo{author}{Duquesne, S.},
  \bibinfo{year}{2017}.
\newblock \bibinfo{title}{{Updating Key Size Estimations for Pairings}}.
\newblock \bibinfo{journal}{Journal of Cryptology} , \bibinfo{pages}{1--39}.
\bibitem[{Birrell and Schneider(2013)}]{birrell2013}
\bibinfo{author}{Birrell, E.}, \bibinfo{author}{Schneider, F.B.},
  \bibinfo{year}{2013}.
\newblock \bibinfo{title}{{Federated Identity Management Systems: A
  Privacy-based Characterization}}.
\newblock \bibinfo{journal}{IEEE Security \& Privacy} \bibinfo{volume}{11},
  \bibinfo{pages}{36--48}.
\bibitem[{Borgia(2014)}]{borgia2014}
\bibinfo{author}{Borgia, E.}, \bibinfo{year}{2014}.
\newblock \bibinfo{title}{The internet of things vision: Key features,
  applications and open issues}.
\newblock \bibinfo{journal}{Computer Communications} \bibinfo{volume}{54},
  \bibinfo{pages}{1--31}.
\bibitem[{Brown et~al.(2009)Brown, Campagna and Vanstone}]{brown2009}
\bibinfo{author}{Brown, D.R.}, \bibinfo{author}{Campagna, M.J.},
  \bibinfo{author}{Vanstone, S.A.}, \bibinfo{year}{2009}.
\newblock \bibinfo{title}{Security of ecqv-certified ecdsa against passive
  adversaries}.
\newblock \bibinfo{journal}{IACR Cryptology ePrint Archive}
  \bibinfo{volume}{2009}, \bibinfo{pages}{620}.
\bibitem[{Brown et~al.(2002)Brown, Gallant and Vanstone}]{ecqv2001}
\bibinfo{author}{Brown, D.R.L.}, \bibinfo{author}{Gallant, R.P.},
  \bibinfo{author}{Vanstone, S.A.}, \bibinfo{year}{2002}.
\newblock \bibinfo{title}{{Provably Secure Implicit Certificate Schemes}}, in:
  \bibinfo{booktitle}{International Conference on Financial Cryptography
  (FC'02)}, pp. \bibinfo{pages}{156--165}.
\bibitem[{Chadwick(2009)}]{chadwick2009a}
\bibinfo{author}{Chadwick, D.}, \bibinfo{year}{2009}.
\newblock \bibinfo{title}{{Federated Identity Management}}.
\newblock \bibinfo{journal}{Foundations of Security Analysis and Design V} ,
  \bibinfo{pages}{96--120}.
\bibitem[{Chadwick and Inman(2009)}]{chadwick2009b}
\bibinfo{author}{Chadwick, D.W.}, \bibinfo{author}{Inman, G.},
  \bibinfo{year}{2009}.
\newblock \bibinfo{title}{{Attribute Aggregation in Federated Identity
  Management}}.
\newblock \bibinfo{journal}{IEEE Computer} \bibinfo{volume}{42},
  \bibinfo{pages}{33--40}.
\bibitem[{Cirani et~al.(2015)Cirani, Picone, Gonizzi, Veltri and
  Ferrari}]{cirani2015iot}
\bibinfo{author}{Cirani, S.}, \bibinfo{author}{Picone, M.},
  \bibinfo{author}{Gonizzi, P.}, \bibinfo{author}{Veltri, L.},
  \bibinfo{author}{Ferrari, G.}, \bibinfo{year}{2015}.
\newblock \bibinfo{title}{{I}o{T}-{O}{A}{S}: An {O}{A}uth-based {A}uthorization
  {S}ervice {A}rchitecture for {S}ecure {S}ervices in {I}o{T} {S}cenarios}.
\newblock \bibinfo{journal}{IEEE Sensors Journal} \bibinfo{volume}{15},
  \bibinfo{pages}{1224--1234}.
\bibitem[{Domenech et~al.(2016)Domenech, Boukerche and Wangham}]{domenech2016}
\bibinfo{author}{Domenech, M.C.}, \bibinfo{author}{Boukerche, A.},
  \bibinfo{author}{Wangham, M.S.}, \bibinfo{year}{2016}.
\newblock \bibinfo{title}{An authentication and {A}uthorization
  {I}nfrastructure for the {W}eb of {T}hings}, in: \bibinfo{booktitle}{ACM
  Symposium on QoS and Security for Wireless and Mobile Networks
  (Q2SWinet'16)}, pp. \bibinfo{pages}{39--46}.
\bibitem[{Fremantle and Aziz(2016)}]{fremantle2016oauthing}
\bibinfo{author}{Fremantle, P.}, \bibinfo{author}{Aziz, B.},
  \bibinfo{year}{2016}.
\newblock \bibinfo{title}{{OA}uthing: {P}rivacy-enhancing {F}ederation for the
  {I}nternet of {T}hings}, in: \bibinfo{booktitle}{Cloudification of the
  Internet of Things (CIoT'16)}, pp. \bibinfo{pages}{1--6}.
\bibitem[{Fremantle et~al.(2014)Fremantle, Aziz, Kopeck\'y and
  Scott}]{fremantle2014}
\bibinfo{author}{Fremantle, P.}, \bibinfo{author}{Aziz, B.},
  \bibinfo{author}{Kopeck\'y, J.}, \bibinfo{author}{Scott, P.},
  \bibinfo{year}{2014}.
\newblock \bibinfo{title}{{Federated Identity and Access Management for the
  Internet of Things}}, in: \bibinfo{booktitle}{International Workshop on
  Secure Internet of Things (SIoT'14)}, pp. \bibinfo{pages}{10--17}.
\bibitem[{Gubbi et~al.(2013)Gubbi, Buyya, Marusic and Palaniswami}]{gubbi2013}
\bibinfo{author}{Gubbi, J.}, \bibinfo{author}{Buyya, R.},
  \bibinfo{author}{Marusic, S.}, \bibinfo{author}{Palaniswami, M.},
  \bibinfo{year}{2013}.
\newblock \bibinfo{title}{Internet of things (iot): A vision, architectural
  elements, and future directions}.
\newblock \bibinfo{journal}{Future generation computer systems}
  \bibinfo{volume}{29}, \bibinfo{pages}{1645--1660}.
\bibitem[{Hammi et~al.(2018)Hammi, Hammi, Bellot and Serhrouchni}]{hammi2018}
\bibinfo{author}{Hammi, M.T.}, \bibinfo{author}{Hammi, B.},
  \bibinfo{author}{Bellot, P.}, \bibinfo{author}{Serhrouchni, A.},
  \bibinfo{year}{2018}.
\newblock \bibinfo{title}{Bubbles of trust: A decentralized blockchain-based
  authentication system for iot}.
\newblock \bibinfo{journal}{Computers \& Security} \bibinfo{volume}{78},
  \bibinfo{pages}{126 -- 142}.
\bibitem[{He and Zeadally(2015)}]{he2015}
\bibinfo{author}{He, D.}, \bibinfo{author}{Zeadally, S.}, \bibinfo{year}{2015}.
\newblock \bibinfo{title}{{An Analysis of RFID Authentication Schemes for
  Internet of Things in Healthcare Environment Using Elliptic Curve
  Cryptography}}.
\newblock \bibinfo{journal}{IEEE Internet of Things Journal}
  \bibinfo{volume}{2}, \bibinfo{pages}{72--83}.
\bibitem[{Hong et~al.(2016)Hong, Levy and Levis}]{Hong16Beetle}
\bibinfo{author}{Hong, J.}, \bibinfo{author}{Levy, A.}, \bibinfo{author}{Levis,
  P.}, \bibinfo{year}{2016}.
\newblock \bibinfo{title}{{Demo: Building Comprehensible Access Control for the
  Internet of Things Using Beetle}}, in: \bibinfo{booktitle}{{ACM International
  Conference on Mobile Systems, Applications, and Services (MobiSys'16)}}.
\bibitem[{Horrow and Sardana(2012)}]{horrow12}
\bibinfo{author}{Horrow, S.}, \bibinfo{author}{Sardana, A.},
  \bibinfo{year}{2012}.
\newblock \bibinfo{title}{{Identity Management Framework for Cloud Based
  Internet of Things}}, in: \bibinfo{booktitle}{International Conference on
  Security of Internet of Things (SecurIT'12)}, pp. \bibinfo{pages}{200--203}.
\bibitem[{Hummen et~al.(2013)Hummen, Ziegeldorf, Shafagh, Raza and
  Wehrle}]{hummen2013towards}
\bibinfo{author}{Hummen, R.}, \bibinfo{author}{Ziegeldorf, J.H.},
  \bibinfo{author}{Shafagh, H.}, \bibinfo{author}{Raza, S.},
  \bibinfo{author}{Wehrle, K.}, \bibinfo{year}{2013}.
\newblock \bibinfo{title}{{Towards Viable Certificate-based Authentication for
  the Internet of Things}}, in: \bibinfo{booktitle}{Workshop on Hot Topics on
  Wireless Network Security and Privacy (HotWiSec'13)},
  \bibinfo{organization}{ACM}. pp. \bibinfo{pages}{37--42}.
\bibitem[{Isaakidis et~al.(2016)Isaakidis, Halpin and Danezis}]{isaakidis2016}
\bibinfo{author}{Isaakidis, M.}, \bibinfo{author}{Halpin, H.},
  \bibinfo{author}{Danezis, G.}, \bibinfo{year}{2016}.
\newblock \bibinfo{title}{{UnlimitID: Privacy-preserving Federated Identity
  Management Using Algebraic MACs}}, in: \bibinfo{booktitle}{Workshop on
  Privacy in the Electronic Society (WPES'16)}, \bibinfo{organization}{ACM}.
  pp. \bibinfo{pages}{139--142}.
\bibitem[{Kim et~al.(2017a)Kim, Kang, Lee and Broman}]{kim2017}
\bibinfo{author}{Kim, H.}, \bibinfo{author}{Kang, E.}, \bibinfo{author}{Lee,
  E.A.}, \bibinfo{author}{Broman, D.}, \bibinfo{year}{2017}a.
\newblock \bibinfo{title}{{A Toolkit for Construction of Authorization Service
  Infrastructure for the Internet of Things}}, in: \bibinfo{booktitle}{ACM/IEEE
  International Conference on Internet-of-Things Design and Implementation
  (IoTDI'17)}, \bibinfo{organization}{ACM/IEEE}. pp. \bibinfo{pages}{147--158}.
\bibitem[{Kim et~al.(2017b)Kim, Hu, Sarkar and Jha}]{kimesiot2017}
\bibinfo{author}{Kim, J.Y.}, \bibinfo{author}{Hu, W.}, \bibinfo{author}{Sarkar,
  D.}, \bibinfo{author}{Jha, S.}, \bibinfo{year}{2017}b.
\newblock \bibinfo{title}{{ESIoT: Enabling Secure Management of the Internet of
  Things}}, in: \bibinfo{booktitle}{ACM Conference on Security and Privacy in
  Wireless and Mobile Networks (WiSec'17)}, \bibinfo{organization}{ACM}. pp.
  \bibinfo{pages}{219--229}.
\bibitem[{Kumari et~al.(2018)Kumari, Karuppiah, Das, Li, Wu and
  Kumar}]{kumari2018}
\bibinfo{author}{Kumari, S.}, \bibinfo{author}{Karuppiah, M.},
  \bibinfo{author}{Das, A.K.}, \bibinfo{author}{Li, X.}, \bibinfo{author}{Wu,
  F.}, \bibinfo{author}{Kumar, N.}, \bibinfo{year}{2018}.
\newblock \bibinfo{title}{A secure authentication scheme based on elliptic
  curve cryptography for iot and cloud servers}.
\newblock \bibinfo{journal}{The Journal of Supercomputing}
  \bibinfo{volume}{74}, \bibinfo{pages}{6428--6453}.
\bibitem[{Li et~al.(2018)Li, Niu, Kumari, Wu, Sangaiah and Choo}]{li2018}
\bibinfo{author}{Li, X.}, \bibinfo{author}{Niu, J.}, \bibinfo{author}{Kumari,
  S.}, \bibinfo{author}{Wu, F.}, \bibinfo{author}{Sangaiah, A.K.},
  \bibinfo{author}{Choo, K.K.R.}, \bibinfo{year}{2018}.
\newblock \bibinfo{title}{{A Three-factor Anonymous Authentication Scheme for
  Wireless Sensor Networks in Internet of Things Environments}}.
\newblock \bibinfo{journal}{Journal of Network and Computer Applications}
  \bibinfo{volume}{103}, \bibinfo{pages}{194--204}.
\bibitem[{Liu et~al.(2012)Liu, Xiao and Chen}]{liu2012}
\bibinfo{author}{Liu, J.}, \bibinfo{author}{Xiao, Y.}, \bibinfo{author}{Chen,
  C.P.}, \bibinfo{year}{2012}.
\newblock \bibinfo{title}{{Authentication and Access Control in the Internet of
  Things}}, in: \bibinfo{booktitle}{International Conference on Distributed
  Computing Systems Workshops (ICDCSW)}, \bibinfo{organization}{IEEE}. pp.
  \bibinfo{pages}{588--592}.
\bibitem[{Malan et~al.(2008)Malan, Welsh and Smith}]{malan08}
\bibinfo{author}{Malan, D.J.}, \bibinfo{author}{Welsh, M.},
  \bibinfo{author}{Smith, M.D.}, \bibinfo{year}{2008}.
\newblock \bibinfo{title}{{Implementing Public-Key Infrastructure for Sensor
  Networks}}.
\newblock \bibinfo{journal}{ACM Transactions on Sensor Networks}
  \bibinfo{volume}{4}, \bibinfo{pages}{22:1--22:23}.
\bibitem[{Maler and Reed(2008)}]{maler2008}
\bibinfo{author}{Maler, E.}, \bibinfo{author}{Reed, D.}, \bibinfo{year}{2008}.
\newblock \bibinfo{title}{{T}he {V}enn of {I}dentity: {O}ptions and {I}ssues in
  {F}ederated {I}dentity {M}anagement}.
\newblock \bibinfo{journal}{IEEE Security \& Privacy} \bibinfo{volume}{6},
  \bibinfo{pages}{16--23}.
\bibitem[{Markmann et~al.(2015)Markmann, Schmidt and
  W{\"a}hlisch}]{markmann2015}
\bibinfo{author}{Markmann, T.}, \bibinfo{author}{Schmidt, T.C.},
  \bibinfo{author}{W{\"a}hlisch, M.}, \bibinfo{year}{2015}.
\newblock \bibinfo{title}{{Federated End-to-end Authentication for the
  Constrained Internet of Things Using IBC and ECC}}.
\newblock \bibinfo{journal}{ACM SIGCOMM Computer Communication Review}
  \bibinfo{volume}{45}, \bibinfo{pages}{603--604}.
\bibitem[{Miettinen et~al.(2016)Miettinen, Huang, Nguyen, Asokan and
  Sadeghi}]{miettinen2016}
\bibinfo{author}{Miettinen, M.}, \bibinfo{author}{Huang, J.},
  \bibinfo{author}{Nguyen, T.D.}, \bibinfo{author}{Asokan, N.},
  \bibinfo{author}{Sadeghi, A.R.}, \bibinfo{year}{2016}.
\newblock \bibinfo{title}{{POSTER: Friend or Foe? Context Authentication for
  Trust Domain Separation in IoT Environments}}, in: \bibinfo{booktitle}{ACM
  Conference on Security \& Privacy in Wireless and Mobile Networks
  (WiSec'16)}, \bibinfo{organization}{ACM}. pp. \bibinfo{pages}{225--226}.
\bibitem[{Neto et~al.(2016)Neto, Souza, Cunha, Nogueira, Nunes, Cotta,
  Gentille, Loureiro, Aranha, Patil et~al.}]{neto2016}
\bibinfo{author}{Neto, A.L.M.}, \bibinfo{author}{Souza, A.L.},
  \bibinfo{author}{Cunha, I.}, \bibinfo{author}{Nogueira, M.},
  \bibinfo{author}{Nunes, I.O.}, \bibinfo{author}{Cotta, L.},
  \bibinfo{author}{Gentille, N.}, \bibinfo{author}{Loureiro, A.A.},
  \bibinfo{author}{Aranha, D.F.}, \bibinfo{author}{Patil, H.K.}, et~al.,
  \bibinfo{year}{2016}.
\newblock \bibinfo{title}{{AoT: Authentication and Access Control for the
  Entire IoT Device Life-cycle}}, in: \bibinfo{booktitle}{ACM Conference on
  Embedded Network Sensor Systems (Sensys'16)}, \bibinfo{organization}{ACM}.
  pp. \bibinfo{pages}{1--15}.
\bibitem[{Ometov et~al.(2019)Ometov, Petrov, Bezzateev, Andreev, Koucheryavy
  and Gerla}]{ometov2019}
\bibinfo{author}{Ometov, A.}, \bibinfo{author}{Petrov, V.},
  \bibinfo{author}{Bezzateev, S.}, \bibinfo{author}{Andreev, S.},
  \bibinfo{author}{Koucheryavy, Y.}, \bibinfo{author}{Gerla, M.},
  \bibinfo{year}{2019}.
\newblock \bibinfo{title}{Challenges of multi-factor authentication for
  securing advanced iot applications}.
\newblock \bibinfo{journal}{IEEE Network} \bibinfo{volume}{33},
  \bibinfo{pages}{82--88}.
\bibitem[{Perrig et~al.(2002)Perrig, Szewczyk, Wen, Culler and
  Tygar}]{perrig02spins}
\bibinfo{author}{Perrig, A.}, \bibinfo{author}{Szewczyk, R.},
  \bibinfo{author}{Wen, V.}, \bibinfo{author}{Culler, D.},
  \bibinfo{author}{Tygar, J.D.}, \bibinfo{year}{2002}.
\newblock \bibinfo{title}{{{SPINS}: Security Protocols for Sensor Networks}}.
\newblock \bibinfo{journal}{Wireless Networks} \bibinfo{volume}{8},
  \bibinfo{pages}{521--534}.
\bibitem[{Porambage et~al.(2014)Porambage, Schmitt, Kumar, Gurtov and
  Ylianttila}]{porambage2014}
\bibinfo{author}{Porambage, P.}, \bibinfo{author}{Schmitt, C.},
  \bibinfo{author}{Kumar, P.}, \bibinfo{author}{Gurtov, A.},
  \bibinfo{author}{Ylianttila, M.}, \bibinfo{year}{2014}.
\newblock \bibinfo{title}{{Two-phase Authentication Protocol for Wireless
  Sensor Networks in Distributed IoT Applications}}, in:
  \bibinfo{booktitle}{IEEE Wireless Communications and Networking Conference
  (WCNC'14)}, \bibinfo{organization}{IEEE}. pp. \bibinfo{pages}{2728--2733}.
\bibitem[{Santos et~al.(2018)Santos, Carneiro, Teixeira, Franco, Henriques and
  Oliveira}]{santos2018}
\bibinfo{author}{Santos, M.L.B.A.}, \bibinfo{author}{Carneiro, J.C.},
  \bibinfo{author}{Teixeira, F.A.}, \bibinfo{author}{Franco, A.M.R.},
  \bibinfo{author}{Henriques, M.A.A.}, \bibinfo{author}{Oliveira, L.B.},
  \bibinfo{year}{2018}.
\newblock \bibinfo{title}{Federated authentication of things: Demo abstract},
  in: \bibinfo{booktitle}{Proceedings of the 17th ACM/IEEE International
  Conference on Information Processing in Sensor Networks},
  \bibinfo{publisher}{IEEE Press}. pp. \bibinfo{pages}{136--137}.
\bibitem[{Shim et~al.(2005)Shim, Bhalla and Pendyala}]{shim2005federated}
\bibinfo{author}{Shim, S.S.Y.}, \bibinfo{author}{Bhalla, G.},
  \bibinfo{author}{Pendyala, V.}, \bibinfo{year}{2005}.
\newblock \bibinfo{title}{{F}ederated {I}dentity {M}anagement}.
\newblock \bibinfo{journal}{IEEE Computer} \bibinfo{volume}{38},
  \bibinfo{pages}{120--122}.
\bibitem[{Stallings(2016)}]{stallings2016cryptography}
\bibinfo{author}{Stallings, W.}, \bibinfo{year}{2016}.
\newblock \bibinfo{title}{{Cryptography and Network Security: Principles and
  Practice}}.
\newblock \bibinfo{publisher}{Pearson}.
\bibitem[{Steiner et~al.(1988)Steiner, Neuman and Schiller}]{steiner1988}
\bibinfo{author}{Steiner, J.G.}, \bibinfo{author}{Neuman, B.C.},
  \bibinfo{author}{Schiller, J.I.}, \bibinfo{year}{1988}.
\newblock \bibinfo{title}{{Kerberos: An Authentication Service for Open Network
  Systems}}, in: \bibinfo{booktitle}{USENIX Winter},
  \bibinfo{organization}{USENIX}. pp. \bibinfo{pages}{191--202}.
\bibitem[{Suh and Devadas(2007)}]{suh2007}
\bibinfo{author}{Suh, G.E.}, \bibinfo{author}{Devadas, S.},
  \bibinfo{year}{2007}.
\newblock \bibinfo{title}{Physical {U}nclonable {F}unctions for {D}evice
  {A}uthentication and {S}ecret {K}ey {G}eneration}, in:
  \bibinfo{booktitle}{ACM/IEEE Design Automation Conference (DAC'07)},
  \bibinfo{organization}{ACM/IEEE}. \bibinfo{publisher}{IEEE}. pp.
  \bibinfo{pages}{9--14}.
\bibitem[{Turkanovi{\'c} et~al.(2014)Turkanovi{\'c}, Brumen and
  H{\"o}lbl}]{turkanovic2014}
\bibinfo{author}{Turkanovi{\'c}, M.}, \bibinfo{author}{Brumen, B.},
  \bibinfo{author}{H{\"o}lbl, M.}, \bibinfo{year}{2014}.
\newblock \bibinfo{title}{{A Novel User Authentication and Key Agreement Scheme
  for Heterogeneous Ad Hoc Wireless Sensor Networks, Based on the Internet of
  Things Notion}}.
\newblock \bibinfo{journal}{Ad Hoc Networks} \bibinfo{volume}{20},
  \bibinfo{pages}{96--112}.
\bibitem[{Ververidis and Polyzos(2008)}]{ververidis2008}
\bibinfo{author}{Ververidis, C.N.}, \bibinfo{author}{Polyzos, G.C.},
  \bibinfo{year}{2008}.
\newblock \bibinfo{title}{Service {D}iscovery for {M}obile {A}d {H}oc
  {N}etworks: a {S}urvey of {I}ssues and {T}echniques}.
\newblock \bibinfo{journal}{IEEE Communications Surveys \& Tutorials}
  \bibinfo{volume}{10}, \bibinfo{pages}{30--45}.
\bibitem[{Wang and Wang(2014)}]{wang2014}
\bibinfo{author}{Wang, D.}, \bibinfo{author}{Wang, P.}, \bibinfo{year}{2014}.
\newblock \bibinfo{title}{{On the Anonymity of Two-factor Authentication
  Schemes for Wireless Sensor Networks: Attacks, Principle and Solutions}}.
\newblock \bibinfo{journal}{Computer Networks} \bibinfo{volume}{73},
  \bibinfo{pages}{41--57}.
\bibitem[{Witkovski et~al.(2015)Witkovski, Santin, Abreu and
  Marynowski}]{witkovski2015}
\bibinfo{author}{Witkovski, A.}, \bibinfo{author}{Santin, A.},
  \bibinfo{author}{Abreu, V.}, \bibinfo{author}{Marynowski, J.},
  \bibinfo{year}{2015}.
\newblock \bibinfo{title}{{An IdM and Key-based Authentication Method for
  Providing Single Sign-on in IoT}}, in: \bibinfo{booktitle}{IEEE Global
  Communications Conference (GLOBECOM'15)}, \bibinfo{organization}{IEEE}. pp.
  \bibinfo{pages}{1--6}.
\bibitem[{Xi et~al.(2016)Xi, Qian, Han, Zhao, Zhong, Li and Zhao}]{xi2016}
\bibinfo{author}{Xi, W.}, \bibinfo{author}{Qian, C.}, \bibinfo{author}{Han,
  J.}, \bibinfo{author}{Zhao, K.}, \bibinfo{author}{Zhong, S.},
  \bibinfo{author}{Li, X.Y.}, \bibinfo{author}{Zhao, J.}, \bibinfo{year}{2016}.
\newblock \bibinfo{title}{{Instant and Robust Authentication and Key Agreement
  Among Mobile Devices}}, in: \bibinfo{booktitle}{ACM Conference on Computer
  and Communications Security (CCS'16)}, \bibinfo{publisher}{ACM}. pp.
  \bibinfo{pages}{616--627}.
\bibitem[{Yavuz(2013)}]{yavuz2013}
\bibinfo{author}{Yavuz, A.A.}, \bibinfo{year}{2013}.
\newblock \bibinfo{title}{{ETA: Efficient and Tiny and Authentication for
  Heterogeneous Wireless Systems}}, in: \bibinfo{booktitle}{ACM Conference on
  Security and Privacy in Wireless and Mobile Networks (WiSec'13)},
  \bibinfo{organization}{ACM}. pp. \bibinfo{pages}{67--72}.
\bibitem[{Zeng et~al.(2018)Zeng, Xu, Zheng, Xiang and Zhou}]{zeng2018}
\bibinfo{author}{Zeng, X.}, \bibinfo{author}{Xu, G.}, \bibinfo{author}{Zheng,
  X.}, \bibinfo{author}{Xiang, Y.}, \bibinfo{author}{Zhou, W.},
  \bibinfo{year}{2018}.
\newblock \bibinfo{title}{E-aua: An efficient anonymous user authentication
  protocol for mobile iot}.
\newblock \bibinfo{journal}{IEEE Internet of Things Journal}
  \bibinfo{volume}{6}, \bibinfo{pages}{1506--1519}.
\bibitem[{Zuo et~al.(2017)Zuo, Zhao and Lin}]{zuo2017}
\bibinfo{author}{Zuo, C.}, \bibinfo{author}{Zhao, Q.}, \bibinfo{author}{Lin,
  Z.}, \bibinfo{year}{2017}.
\newblock \bibinfo{title}{{Authscope: Towards Automatic Discovery of Vulnerable
  Authorizations in Online Services}}, in: \bibinfo{booktitle}{ACM Conference
  on Computer and Communications Security (CCS'17)},
  \bibinfo{organization}{ACM}. pp. \bibinfo{pages}{799--813}.

\end{thebibliography}





\end{document}